\documentclass[format=acmsmall, review=False, screen=true]{acmart}

\usepackage{booktabs} 

\usepackage[ruled]{algorithm2e} 

\SetAlFnt{\small}
\SetAlCapFnt{\small}
\SetAlCapNameFnt{\small}
\SetAlCapHSkip{0pt}
\IncMargin{-\parindent}

\usepackage{amsmath}
\usepackage{graphicx}
\usepackage{url}
\usepackage{amsfonts,amssymb}
\usepackage{booktabs}
\usepackage{multirow}
\usepackage{lscape}

\setcopyright{acmlicensed}

\begin{document}
\title{Attentive Aspect Modeling for Review-aware Recommendation}

\author{Xinyu Guan}
\affiliation{
\institution{Xi'an Jiaotong University}
\department{Systems Engineering Institute}
\streetaddress{28, Xianning West Road}
\city{Xi'an}
\postcode{710049}
\country{P. R. China}
}
\email{xinyu_guan@foxmail.com}

\author{Zhiyong Cheng$^{\dag}$}
\affiliation{
  \institution{ Qilu University of Technology (Shandong Academy of Sciences)}
  \department{Shandong Computer Science Center (National Supercomputer Center in Jinan), Shandong Artificial Intelligence Institute}
  \streetaddress{19 Keyuan Road}
     \city{Jinan}
    \state{Shandong}
    \postcode{250014}
  \country{China}
 }
\email{jason.zy.cheng@gmail.com}

\author{Xiangnan He}
\affiliation{
\institution{University of Science and Technology of China}
\department{School of Information Science and Technology}
\streetaddress{443 Huangshan Road}
\city{Hefei}
\postcode{230031}
\country{China}
}
\email{xiangnanhe@gmail.com}

\author{Yongfeng Zhang}
\affiliation{
\institution{Rutgers University}
\department{Department of Computer Science}
\streetaddress{110 Frelinghuysen Road}
\city{Piscataway}
\postcode{08854}
\country{USA}
}
\email{zhangyf07@gmail.com}

\author{Zhibo Zhu}
\affiliation{
\institution{Xi'an Jiaotong University}
\department{Systems Engineering Institute}
\streetaddress{28 Xianning West Road}
\city{Xi'an}
\postcode{710049}
\country{P. R. China}
}
\email{zdh96.zzb@stu.xjtu.edu.cn}

\author{Qinke Peng}
\affiliation{
\institution{Xi'an Jiaotong University}
\department{Systems Engineering Institute}
\streetaddress{28, Xianning West Road}
\city{Xi'an}
\postcode{710049}
\country{P. R. China}
}
\email{qkpeng@mail.xjtu.edu.cn}

\author{Tat-Seng Chua}
\affiliation{
\institution{National University of Singapore}
\streetaddress{13 Computing Drive}
\department{School of Computing}
\city{Singapore}
\postcode{117417}
\country{Singapore}
}
\email{dcscts@nus.edu.sg}

\thanks{$^{\dag}$ Corresponding Author. \\ This work was finished when Xinyu Guan was a visiting student at the National University of Singapore. The support provided by China Scholarship Council (CSC) during the visit of Xinyu Guan to National University of Singapore is acknowledged. The first author claims that this work is under the supervision of Dr. Zhiyong Cheng and Dr. Xiangnan He. This work is supported by the National Natural Science Foundation of China (grant numbers 61872288). This work is also supported by the NExT research centre, which is supported by the National Research Foundation, Prime Minister's Office, Singapore under its IRC@SG Funding Initiative.
}

\begin{abstract}
In recent years, many studies extract aspects from user reviews and integrate them with ratings for improving the recommendation performance. The common aspects mentioned in a user's reviews and a product's reviews indicate indirect connections between the user and product. However, these aspect-based methods suffer from two problems. First, the common aspects are usually very sparse, which is caused by the sparsity of user-product interactions and the diversity of individual users' vocabularies. Second, a user's interests on aspects could be different with respect to different products, which are usually assumed to be static in existing methods. In this paper, we propose an Attentive Aspect-based Recommendation Model (AARM) to tackle these challenges. For the first problem, to enrich the aspect connections between user and product, besides common aspects, AARM also models the interactions between synonymous and similar aspects. For the second problem, a neural attention network which simultaneously considers user, product and aspect information is constructed to capture a user's attention towards aspects when examining different products. Extensive quantitative and qualitative experiments show that AARM can effectively alleviate the two aforementioned problems and significantly outperforms several state-of-the-art recommendation methods on top-N recommendation task.
\end{abstract}



\setcopyright{acmlicensed}
\acmJournal{TOIS}
\acmYear{2019} \acmVolume{1} \acmNumber{1} \acmArticle{1} \acmMonth{1} \acmPrice{15.00}\acmDOI{10.1145/3309546}

\maketitle

\renewcommand{\shortauthors}{X. Guan et al.}

\section{Introduction}
Recommender systems help users find their potentially interested products from an enormous list of products. Matrix Factorization (MF) methods \cite{5197422} are widely adopted in recommendation systems because of its accuracy and scalability. MF methods usually rely on the explicit (e.g., user ratings) or implicit (e.g., click behaviors) interactions between users and products for recommendation. However, a rating or binary interaction can only reflect the user's overall attitude towards a product, which does not include information about the underlying reasons for the user behavior. As a result, it is difficult for MF methods to model user's fine-grained preferences on specific product features and provide explanation to recommendations. 

To tackle these limitations, researches have attempted to utilize reviews to alleviate the data sparsity problem and provide more explainable recommendations \cite{Chen2015,cheng2018ijcai,He:2015:TRE:2806416.2806504,CaoDaTois,cheng2019mmalfm}. As accompanying information of ratings, the textual review expresses user's opinions on different product features, and thus contains more fine-grained information about the user preference. Different strategies have been applied to incorporate reviews into MF models, including sentiment analysis \cite{opinion_driven_mf},  representation learning \cite{zhang2017joint,Catherine:2017:TLT:3109859.3109878}, and topic models \cite{mcauley2013hidden,rblt2016ijcai}. Although these methods have achieved some progress, the generated vector representations of users and products are still latent and thus cannot explicitly model user's preference on specific product features, which could impede their performance.

Another direction is to leverage the aspects mentioned in user reviews for recommendation. In this paper, \textbf{aspect} is defined as the words or phrases used by users in their product reviews to describe product features. For example, ``battery life'' and ``battery duration'' are two different aspects while they refer to the same product feature.
There are already some methods which detect aspects in user reviews and leverage them to model user's fine-grained preferences to specific product features \cite{Zhang:2014:EFM:2600428.2609579,ijcai2017-674}. 
For example, EFM \cite{Zhang:2014:EFM:2600428.2609579} conducted aspect-level sentiment analysis to extract user's preference and product's quality on specific product feature, then incorporated the results into an MF framework to provide more accurate recommendation. SULM \cite{Bauman:2017:ABR:3097983.3098170} and LRPPM \cite{Chen2016LRF} went beyond EFM \cite{Zhang:2014:EFM:2600428.2609579} by using more effective methods to identify the impact of each aspect on the overall rating.
However, these methods rely highly on the accuracy of external sentiment analysis tools.

Besides the above mentioned limitations, these methods also suffers from the following two problems. 
First, for each user-product pair, they only consider the shared aspects in the user's reviews and the product's reviews. However, due to the sparsity of user-product interactions and users' diverse language usages, the number of common aspects mentioned in the reviews of both the targeted user and product is usually very limited. Second, a user's concerned aspects may be different for different products (even in the same category). For example, a user may mostly concern about ``special effects'' when watching a super-hero movie, while pay more attentions to the ``plot'' for a suspense movie.

Motivated by the above concerns, in this paper, we propose an Attentive Aspect-based Recommendation Model (AARM), which can effectively tackle the above two problems. For the first problem of \emph{aspect sparsity}, AARM models the interactions between synonymous and similar aspects to alleviate it, where \emph{synonymous aspects} are the ones referring to the same product feature (e.g., ``storyline'' and ``plot''); and \emph{similar aspects} are those of different features that are closely related (e.g., ``battery life'' and ``charging speed''). Intuitively,  a user's attention to an unmentioned aspect can be inferred through its similar aspects. For instance, a user who cares about ``battery life'' of cellphones may also care about its ``charging speed'', although ``charging speed'' has never been mentioned in this user's reviews. In our model, an aspect extracted from reviews is first represented as an embedding vector. Then a user $u$'s satisfaction about product $v$ according to aspect $a$ is estimated by calculating the interactions between $a$ and all the aspects mentioned in $v$'s reviews. And an attention module is designed to pick up the interactions between meaningful aspect pairs. In this way, we achieve the goal of capturing the interactions between synonymous and similar aspects.

For the second problem of \emph{identifying user's varied interests on aspects}, AARM introduces another attention module which takes user, product and aspect information into consideration. In this way, user's varied interests on aspects can be captured by the product-dependent user attention. Instead of rating prediction, we target the top-N recommendation task with a pair-wise learning-to-rank method, which is the most practically used recommendation scenario in real-world systems \cite{Cremonesi:2010:PRA:1864708.1864721, TOIS2016RANK}. To this end, our model estimates a user $u$'s satisfaction towards an product $v$ by (1) estimating $v$'s performances on $u$'s concerned aspects; and (2) identifying the impacts of these aspects on the overall satisfaction.

We evaluate our model on five product datasets from Amazon on the top-N recommendation task. Experimental results show that AARM outperforms several state-of-the-art methods. Comparative experiments have also been conducted to demonstrate the importance of modeling interactions between different aspects and the effectiveness of our attention module on capturing user's varied attentions towards aspects. 
Our main contributions are outlined as follows.
\begin{itemize}
\item We propose a novel recommendation method to model the interactions between both the same and the different aspects, which helps to alleviate the aspect sparsity problem in reviews. To the best of our knowledge, this is the first attempt to model the interactions between different aspects to model user preferences in recommendation. And the method to capture the similarity relation between different aspects can also be used in other recommendation scenes (e.g. recommendation with tags or item metadata).
\item We design an attention mechanism in AARM to capture user's varied attentions on different aspects towards various products. The careful design of the inputs and structure of this attention module has been demonstrated to be very effective on improving the recommendation accuracy in the experiments.
\item We conduct extensive experiments on real-world datasets to demonstrate the effectiveness of our model. Experimental results show that our method can achieve superior performance by a large margin.
\end{itemize}

The reminder of the paper is organized as follows. We first discuss existing works related to our method in Section 2. In Section 3, we describe the details of AARM and describe how to train the model. In Section 4, we describe the experimental settings and report the results to verify our assumptions and compare our methods with some state-of-the-art baselines. Finally, in Section 5, we conclude the paper.

\section{Related Work}
In recent years, many researchers have paid more attentions to users' product reviews in order to improve the recommendation accuracy and provide recommendation explanation. According to how these methods utilize user reviews, we broadly group them into three categories: \textit{review-level}, \textit{topic-level} and \textit{aspect-level} methods. In this section, we first review these three types of review-based methods, and then briefly discuss the recommendation methods with attention mechanism which is an important component in our model.

\subsection{Review-level Methods}
Review-level methods treat the review as a single piece of information and incorporate it with ratings. The opinion-driven matrix factorization model \cite{opinion_driven_mf} calculates the overall opinion of a review by summing up the orientations of opinion words in the text, and then combines it with numerical ratings for rating prediction. Meng et al. \cite{MIRROR2018AAAI} incorporated other users' emotions towards a review to calculate the importance of this review in the training of matrix factorization model. Some methods concatenate all the reviews belonging to a user (or item) as a user (or item) document, and then employ deep learning methods to learn the continuous vector representation for the user (or item) \cite{Catherine:2017:TLT:3109859.3109878,Zheng:2017:JDM:3018661.3018665,guo2018mm,zhang2017joint}. For example, Transnets \cite{Catherine:2017:TLT:3109859.3109878} and DeepCoNN \cite{Zheng:2017:JDM:3018661.3018665} process the user and item documents with convolutional neural network to generate the vector representation for users and items. JRL \cite{zhang2017joint} adopts the PV-DBOW model \cite{Le:2014:DRS:3044805.3045025}, which is an unsupervised methods to learn the continuous vector representations for documents, and the user and item vector representations from their reviews. In Transnets, DeepCoNN and JRL, in order to estimate the matching degree between a user and an item, reviews of the user or item are compressed to a vector which is an overall representation of the reviews. In this way, these review-level methods neglect the user-item interactions at the review components (e.g. the user's opinions about the product's specific features) level, which can be used to connect the user with candidate products and provide more explainable recommendation.

\subsection{Topic-level Methods}
Topic-level methods build probabilistic graphical model to extract topics from reviews. HFT \cite{mcauley2013hidden} combines topic vectors from reviews with latent factors from ratings to improve rating prediction accuracy. Subsequently, some studies employ different topic models and combination strategies for the review-based rating prediction task. For example, different from HFT, ITLFM \cite{ITLFM2016} linearly combines the latent topics and the latent factors. CMR \cite{CMR2014} is a probabilistic graphical model which simultaneously associates the review text, the hidden user communities and item group relationship with numerical ratings. RBLT \cite{rblt2016ijcai} also utilizes LDA to extract topics from review text. Then the preference distribution vector of each user and the recommendability distribution vector of each item are combined with vanilla matrix factorization model for rating prediction. More recently, Cheng et al. \cite{Cheng2018www} defined a high-level semantic concept `aspect' as a probability distribution of topics. They proposed the ATM model to extract topics from reviews and associated the topics with `aspects', and then proposed the ALFM model to associate latent factors with `aspects'. In this way, topics are correlated with factors via the `aspects' indirectly. To estimate the overall rating score, they first calculated the item's scores on each aspects and then summed them up using aspect importance as weights. Similarly, MMALFM \cite{cheng2019mmalfm} follows the definition of `aspect' in \cite{Cheng2018www} and jointly models the `aspects' in textual reviews and item images. These topic-level methods usually focus on rating prediction task, while we are targeting at top-N recommendation. Similar to review-level methods, when estimating the matching degree between a user and a product, these topic-level methods also neglect the interactions between the components of the user and the product's reviews. And it is difficult to associate a topic, which is a probabilistic distribution over words or phrases, with specific product features. Because of these limitations, these methods are incapable of capturing user's preference towards product features in a finer-grained manner and thus provide more accurate and explainable recommendations.

\subsection{Aspect-level Methods}
Aspect-level methods extract aspects from reviews and incorporate them with ratings for recommendation. The proposed method in this paper falls into this category. Ganu et al. \cite{Ganu:2013:IQP:2379838.2379890} manually defined six aspects and four sentiments for restaurant reviews and used a regression-based method for rating prediction. Zhang et al. \cite{Zhang:2014:EFM:2600428.2609579} employed an unsupervised tool for aspect extraction and aspect-level sentiment analysis. Aspect and sentiment outputs from this step were integrated with matrix factorization methods for rating prediction. Chen et al. \cite{Chen2016LRF} proposed a tensor-matrix factorization method to select the most interesting product features for each user with a learning to rank method. The rating scores were then predicted as the weighted summation of the product's sentiment scores on the user's most cared product features. Bauman et al. \cite{Bauman:2017:ABR:3097983.3098170} also extracted aspects and conducted aspect-level sentiment analysis with external tools. The results of aspect-level sentiment analysis were used in their model SULM as the ground-truth labels to train a latent factor model for every aspect. These aspect-level latent factor models were then used to predict user's sentiment scores toward each aspect of a product. The number of parameters in SULM is very large as a user or product usually has many aspects. As we can see, the above methods often rely on external sentiment tools for aspect-level analysis. 

Specifically, there are also some papers which pay attention to users' varied interests. Chen et al. \cite{Chen2016LRF} proposed an aspect ranking method to capture user's varied interests while they paid more attention to a user's interest variation over different categories. A$^3$NCF \cite{cheng2018ijcai}, ALFM \cite{Cheng2018www} and ANR \cite{chin2018anr} also try to capture users' varied interests towards aspects. Specially, A$^3$NCF and ANR also use neural attention layers to do it. But there are some important differences between them and our method. First, the `aspect' defined in A$^3$NCF, ALFM and ANR are different from the one defined in our model. In A$^3$NCF, `aspect' is defined as a combination of topic vector and embedding vector. In ALFM, `aspect' is defined as a probability distribution of topics and thus ALFM is more like a topic-level model. In ANR, an `aspect' of a user is a weighted sum of all the words' embeddings in the user's reviews. Different from them, `aspects' in our model are words or phrases directly extracted from reviews which are much more fine-grained concept. Second, A$^3$NCF, ALFM and ANR have not considered interactions between different aspects. Different from them, our method models these interactions because intuitively these aspects are not independent of each other. Third, those three existing methods are originally designed for rating prediction, while our model is designed for top-N recommendation.

He et al. \cite{He:2015:TRE:2806416.2806504} did not conduct sentiment analysis but adopted the aspect frequency information in reviews to construct the user-item-aspect tripartite graph for recommendation.
The improved performance in \cite{He:2015:TRE:2806416.2806504} from baselines verified that the aspect mention signals in reviews could have already been able to reflect user's interests on aspects. Similarly, in AARM we do not conduct sentiment analysis on reviews explicitly, which helps to simplify the model design and implementation. Moreover, AARM considers both the interactions between different aspects and the user's varied preference towards aspects, which are neglected by previous studies.

\subsection{Attention Mechanism}
In recent years, many deep learning-based recommendation methods have been proposed and achieved good performance in many tasks \cite{NFM2017,he2017neural,TOIS2017YUM,Tois2018Quote}. The attention mechanism which can assign adaptive weights for a set of features has also been employed in recommendation models \cite{chen2017attentive, he2018nais, CaoDaAGR, Ebesu2018CMN, wsdm2019chen}. For example, in the NARRE model \cite{NARRE2018WWW} for review-based rating prediction, Chen et al. introduced an attention module to calculate the usefulness of reviews. In TEM \cite{Wang2018TEM} which utilizes user and item's side information for explainable recommendation, neural attention layer is used to assign weights to cross features and provide recommendation explanation. ACF \cite{chen2017attentive}, which focuses on multimedia recommendation, uses a component-level attention module to find informative components for multimedia items (images/videos), and a item-level attention module to select representative items to to represent users' preferences. AFM \cite{AFM2017}, which is an extension of FM machine \cite{NFM2017,rendle2010fm}, uses an attention neural network to discriminate the importance of different feature interactions. A$^3$NCF \cite{cheng2018ijcai} and ANR \cite{chin2018anr} also have attention modules which have been discussed in last section. Compared with these methods, we specially design two attention modules for the fine-grained modeling of product features extracted from user reviews. The user-level attention module in AARM is built to find out the user's most concerned product features for a candidate product, while the aspect-level attention module is constructed to select informative aspect interactions.

\section{Attentive Aspect-based Recommendation Model}

\begin{figure}[t]
\centering
\scalebox{0.6}{
\includegraphics[width = 1.0\textwidth]{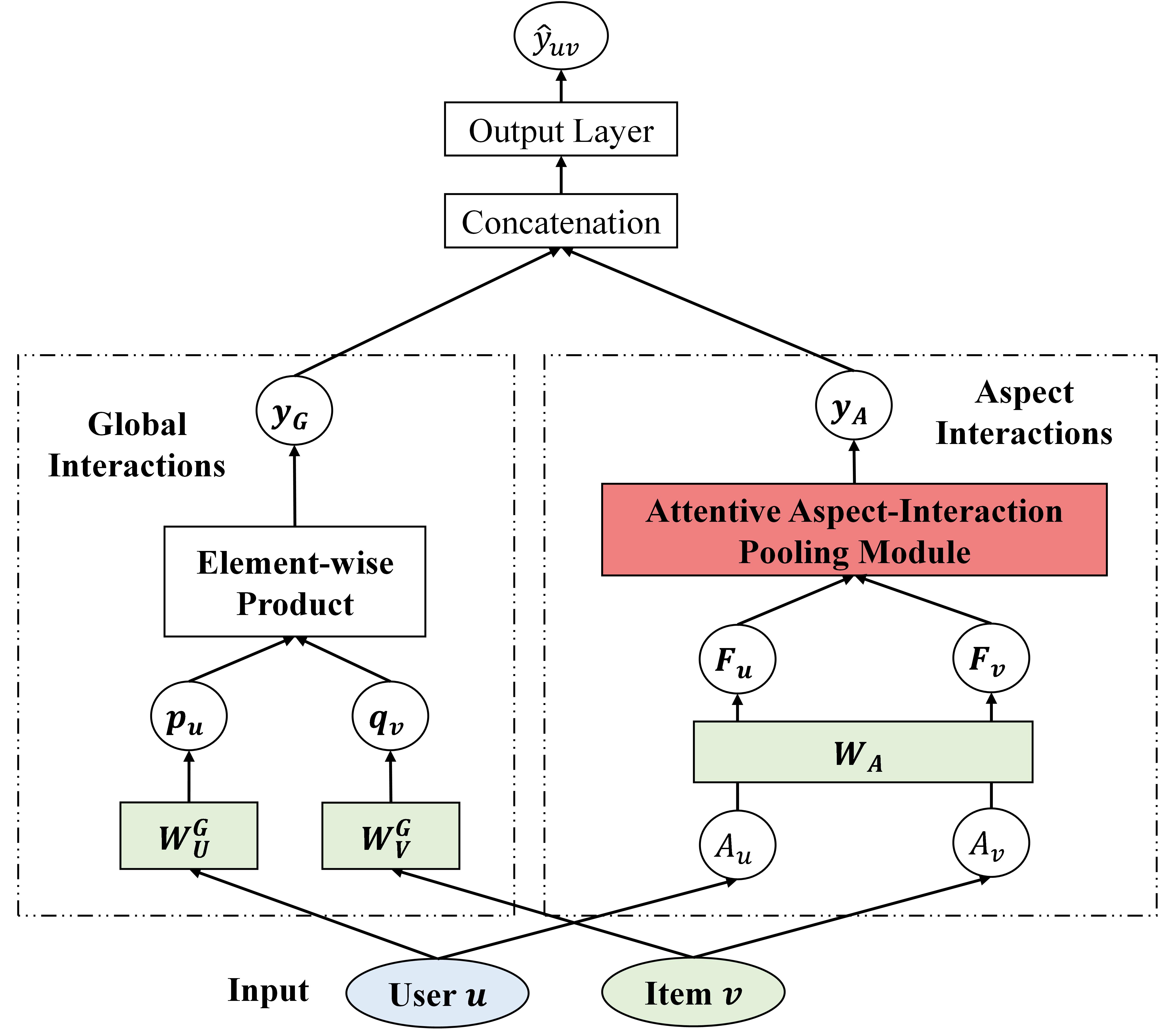}
}
\caption{\textbf{Attentive Aspect-based Recommendation Model}}
\end{figure}

In this section, we first provide an overview of our method and define some important notations, and then introduce how to extract aspects from user reviews. After that, we describe the structure and details of the proposed AARM model. 
In particular, we elaborate how AARM could model the interactions between different aspects and handle user's varied interests in aspects. Finally, we discuss the parameter inference in AARM.

\subsection{Preliminaries}
Given a user set $U = \{ {u_1},{u_2},...{u_{|U|}}\}$ and a product set $V = \{ {v_1},{v_2},....{v_{|V|}}\}$, AARM estimates a satisfaction score $\hat{y}_{uv}$ for an user $u$ towards a product $v$. The candidate products are then ranked in a descending order of $\hat{y}$ and the top N products are recommended to $u$. In our method, aspects extracted from user reviews are used as the explicit features of users and products. We define $A = \{ {a_1},{a_2},....{a_{|A|}}\}$ as the aspect set of the dataset. The aspects that have been mentioned in the reviews of user $u$ is represented as $A_u$, which is a subset of $A$. Similarly, product $v$'s aspects that have been mentioned in $v$'s reviews are represented as $A_v$. Product $v$'s rating given by user $u$ is denoted as $r_{uv}\in R$, where $R$ is the collection of ratings.

The structure of AARM is shown in Figure 1. In the input layer, users and products are represented as binarized sparse vectors using the one-hot encoding method. Above the input layer, the Aspect Interactions part is used to model the interactions between the aspects from user $u$'s aspect set $A_u$ and the aspects from the product $v$'s aspect set $A_v$. Because a user's review for a product may not cover all the factors which can influence the user's satisfaction towards the product, the aspects extracted from review text may not be able to fully explain the rating. Hence the Global Interactions part is stacked above the input layer to model the implicit factors which influence user's decision but have not been discussed in the reviews. Finally, the results of aforementioned two parts are concatenated as the input to the Output Layer.

\begin{figure*}
\begin{center}
  \includegraphics[width=1\textwidth]{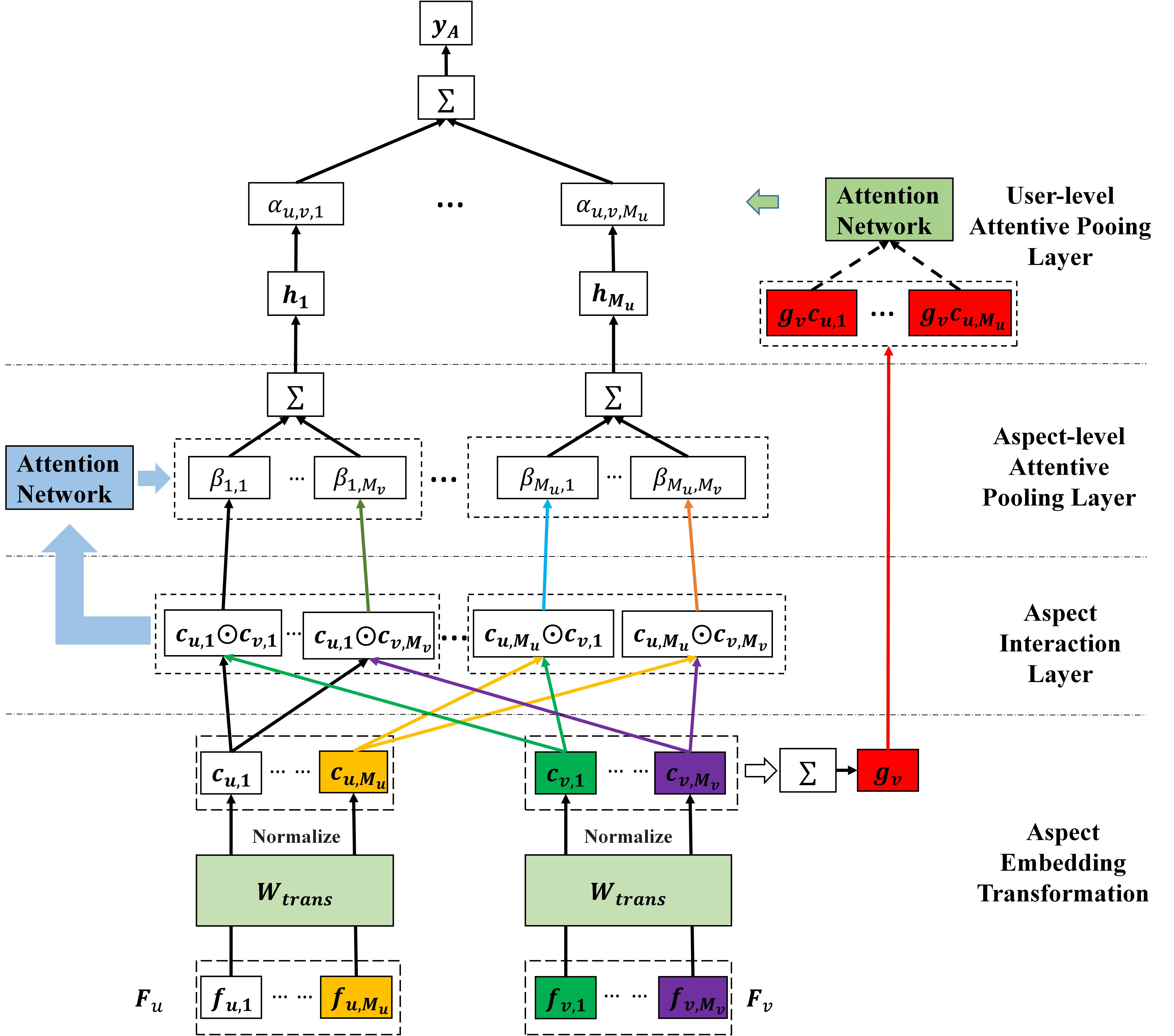}
  \caption{\textbf{The Attentive Aspect-Interaction Pooling Module.}}
\end{center}
\label{fig:2}
\end{figure*}

\subsection{Aspect Interactions Part}
In the Aspect Interactions part, given a user $u$ and a product $v$, aspects are first extracted from their reviews and used to construct their aspect sets $A_u$ and $A_v$, respectively. To model the similarity between aspects, instead of one-hot encoding or bag-of-words model, embedding layers are used in AARM to represents aspects as continuous vectors. Specifically, aspect embedding matrix $\textbf{W}_A\in \mathbb{R}^{d_a\times |A|}$ is defined to project aspects from $A_u$ and $A_v$ to $\textbf{F}_u\in \mathbb{R}^{d_a\times M_u}$ and $\textbf{F}_v\in \mathbb{R}^{d_a\times M_v}$, respectively, where $d_a$ is the dimension of aspect embeddings, and $M_u$ and $M_v$ are respectively the number of aspects in $A_u$ and $A_v$. The $i$th aspect in $A_u$ is projected to $\textbf{f}_{u,i}$ which is the $i$th column of $\textbf{F}_u$. Similarly, aspects in $A_v$ are projected to the embedding vectors in $\textbf{F}_v$. Next, Attentive Aspect-Interaction Pooling Module is designed to model the bi-interactions between the aspect embeddings of $\textbf{F}_u$ and that of $\textbf{F}_v$, and outputs a vector $\textbf{y}_A$ to represent the preference information in user reviews.

\subsubsection{Aspect Extraction}
Because the main contribution of this paper focuses on how to leverage aspects for personalized recommendation, we refer to external tools for aspect extractions. In this paper, we use the Sentires\footnote{http://yongfeng.me/software/}, which has been successfully used in \cite{Zhang:2014:EFM:2600428.2609579,zhang2014users} for aspect extraction. Other aspect extraction tools can also be applied. This toolkit extracts aspects via a hybrid of rule-based and machine learning algorithms. Given a dataset, it generates an aspect lexicon, which is used to build the aspect set $A$ of the dataset in this paper. With this toolkit, we could obtain user aspect set $A_u$ for each user $u \in U$, and product aspect set $A_v$ for each product $v \in V$ by extracting the mentioned aspects from their reviews. Some examples of the automatically extracted aspects are shown in Table 3.

Note that the size of aspect set varies for different users or products. To accelerate the training of AARM, we pad all the user aspect set into the same length $M_u$ and pad all the product aspect set into the same length $M_{v}$. Taking user aspect set as example, we define a meaningless aspect $<PAD>$ and add it to the end of user aspect sets whose lengths are less than the predefined size $M_{u}$. For $A_u$ whose length is larger than $M_{u}$, we calculate the \emph{TF-IDF} score \cite{salton1975vector} of each $a\in A_u$, and truncate $A_u$ to $M_u$ aspects by dropping the aspects with low \emph{TF-IDF} scores. The \emph{TF-IDF} score is defined as:
\begin{equation}
tfidf_u(a) = \frac{tf_u(a)}{\sum\limits_{i \in A_u}{tf_u(i)}}\cdot ln\frac{|U|}{df(a)+1}
\end{equation}
where $tf_u(a)$ is the frequency of $a$'s occurrence in $u$'s reviews, $|U|$ is the number of users, and $df(a)$ is the number of users who mentioned $a$.
All the product aspect sets are padded into the same length $M_v$ in a similar way.

\subsubsection{Attentive Aspect-Interaction Pooling Module}
As shown in Figure 2, given $\textbf{F}_u$ and $\textbf{F}_v$ as input, there are four parts in this module: \textbf{\textit{aspect embedding transformation}}, \textbf{\textit{aspect interaction layer}}, \textbf{\textit{aspect-level attentive pooling layer}}, and \textbf{\textit{user-level attentive pooling layer}}. The final output of this module is the vector $\textbf{y}_A(u,v)$ which represents the overall satisfaction of a user $u$ towards a product $v$ estimated with review text. In this module, we hold the assumption that $u$'s overall satisfaction for $v$ is based on $v$'s performances on $u$'s concerned aspects (i.e. aspects from $A_u$). This module works as follows. First, for each aspect $a\in A_u$, the aspect interaction layer and aspect-level attentive pooling layer are employed to estimate $v$'s performance on $a$, where the performance is represented as vector $\textbf{h}_a(u,v)$. Then the user-level attentive pooling layer is used to estimate $u$'s preference towards $v$ by integrating $\textbf{h}_a(u,v)$ for all the aspect $a\in A_u$ and represent the preference as a vector $\textbf{y}_A(u,v)$. Finally, $\textbf{y}_A(u,v)$ will be combined with the result of Global Interaction part and further input into the output layer to estimate the user $u$'s satisfaction score towards the product $v$.

\paragraph{Aspect Embedding Transformation}

To model the interactions between synonymous and related aspects, we expect the vector representation of aspect to encode the similarity relation between aspects. In this paper, the Word2vec model \cite{word2vec}, which is able to encode many linguistic regularities and patterns, is used to pre-train aspect embeddings with the review texts in each dataset. The aspect embedding matrix $\textbf{W}_A$ is initialized with the pre-trained embeddings and its parameters would not be tuned during the training of AARM. Instead, a trainable matrix $\textbf{W}_{trans}\in \mathbb{R}^{d_a\times d_a}$ is defined to customize the pre-trained aspect embedding $\textbf{f}$ (column vector in $\textbf{F}_u$ or $\textbf{F}_v$) to make it oriented towards our recommendation task. Then these customized embeddings are normalized as:
\begin{equation}
\textbf{c} = \frac{\textbf{W}_{trans}\textbf{f}}{\|\textbf{W}_{trans}\textbf{f}\|}.
\end{equation}
Here $\|\textbf{x}\|$ is the Euclidean norm of $\textbf{x}$. In this paper, aspect interaction between two aspects is defined as the element-wise product of their embedding vectors. By normalizing the aspect embeddings with their corresponding Euclidean norms, the calculation of interaction between two aspects is similar to calculating their cosine similarity. As illustrated in \cite{word2vec}, if two words have higher semantics and syntax similarities, their embeddings generated by Word2vec would have larger cosine similarity. In this way, the results of aspect interactions are associated with the semantics and syntax relations between aspects, which helps in identifying the synonymous and related aspects. Alternatively, we can also directly tune the aspect embedding matrix $\textbf{W}_A$ during the training of AARM for top-N recommendation. We will compare the performances of these different settings in the experiment section.

\paragraph{Aspect Interaction Layer}

This layer maps the vector representations of aspects in $A_u$ and $A_v$ to a set of $d_a$-dimensional interacted vectors. The aspect interaction between aspect $i\in A_u$ and $j\in A_v$ is defined as the element-wise product of their embedding vector $\textbf{c}_i$ and $\textbf{c}_j$. Hence the output of the aspect interaction layer can be represented as a set of vectors:
\begin{equation}
f_{AI}(u,v)=\{{\textbf{c}_i}\odot {\textbf{c}_j}(x_ix_j)\}_{i\in A_u, j\in A_v}.
\end{equation}
Here $x_i\in \{0,1\}$ is the masking indicator, where $x_i=0$ if $i$ is the meaningless aspect $<PAD>$ (defined for padding). To implement the masking operation in AARM, we define an aspect masking vector $\textbf{W}_{mask}\in \mathbb{R}^{d_a\times |A|}$, where the column of aspect $<PAD>$ is a zero vector, and the columns of other aspects in $A$ are vectors of ones. Before calculating the interactions between the aspect $i\in A_u$ and aspects in $A_v$, we first calculate the element-wise product between $\textbf{c}_i$ and its corresponding column in $\textbf{W}_{mask}$. After the masking operation, the embedding vector of aspect $<PAD>$ is transformed into a zero vector. In this way, we make sure that the interactions between aspect $<PAD>$ and other aspects are zero vectors. As shown in the following sections, these zero vectors would not influence AARM's final predictions.

As shown in Equation (3), besides the same aspects, the interactions between different aspects (when $i\ne j$) are also calculated. This is because we want to model the interactions between synonymous and similar aspects to alleviate the problem that \textit{the same aspects shared in a user's reviews and a product's reviews are usually very sparse}. However, interactions between unrelated aspects are also considered in Equation (3). To emphasize on interactions between related aspects and filter out noisy interactions, the aspect-level attentive pooling layer is stacked above this layer.

\paragraph{Aspect-level Attentive Pooling Layer}

In the aspect interaction layer, for each aspect $a\in A_u$, we calculate its interaction with all the aspects in $A_v$. Intuitively, some aspect interactions should be given more attention than others. For example, the interactions between the same, synonymous or similar aspects usually contain more information about the product's performance on the user's concerned aspects. Hence an attention module is designed to focus on important aspect interactions. Word2vec embeddings of similar words would have higher cosine similarities \cite{word2vec}. Inspired by this, for aspect pair $i$ and $j$, the input of attention layer is defined as the element-wise product of their normalized embedding vector $\textbf{c}_i$ and $\textbf{c}_j$ to mimic the cosine similarity between their embedding vectors. And the aspect-level attention layer is defined as:
\begin{equation}
\begin{array}{l}
{{\hat \beta }_{i,j}} = {\textbf{w}_{att_1}}^T({\textbf{c}_i}\odot {\textbf{c}_j})(x_ix_j),\\
{\beta _{i,j}} = \frac{{{\rm{exp}}\left( {{{\hat \beta }_{i,j}}} \right)}}{\sum\limits_{y \in {A_v}} {{\rm{exp}}\left( {{{\hat \beta }_{i,y}}} \right)}}.
\end{array}
\end{equation}
Here $\textbf{w}_{att_1}\in \mathbb{R}^{d_{a}}$  is a learnable vector, and $\beta_{i,j}$ is the attention value of the interaction between aspect $i$ and $j$.

To estimate the product's performance on the user's aspect $a_i\in A_u$, we compress all the interactions between $a_i$ and aspects in $A_v$ with a weighted sum pooling where $\beta$ is used as the weight:
\begin{equation}
\textbf{h}_i = \sum\limits_{j \in {A_v}} {{\beta_{i,j}}({\textbf{c}_i}\odot {\textbf{c}_j})(x_ix_j)}.
\end{equation}
The output of this layer is the vector set $\{\textbf{h}_i \in \mathbb{R}^{d_a}\}_{a_i\in A_u}$.

\paragraph{User-level Attentive Pooling Layer}

We can integrate the vector set $\{\textbf{h}_i\}_{a_i\in A_u}$, which represents how the product fits the user's requirements on each aspect, and thus to produce estimation of the user's overall satisfactory on this product. Intuitively, different users may focus on different aspects even when purchasing the same products. For example, when purchasing a cell phone, some users are more concerned about battery duration while some other users are more concerned about the performance of CPU. Furthermore, when purchasing different products, a user's most concerned product features may be different. In other words, a user's attention towards a aspect when purchasing a specific product is influenced by the characteristics of the user, the aspect and the product simultaneously.

To estimate user $u$'s interest towards aspect $a\in A_u$ when purchasing a specific product $v$, a user-level attentive pooling layer is designed in AARM. The input of this attention layer should contain not only information of current aspect $a$, but also information of product $v$. Intuitively, if an aspect $a\in A_u$ is more important to product $v$, the user should pay more attention to the aspect $a$ as compared with other unrelated aspects in $A_u$. The importance of the user's aspect $a$ with respect to a product $v$ can be measured by the similarities between $a$ and the aspects that has been mentioned in $v$'s reviews (i.e., aspects from $A_v$). To calculate the importance of aspect $a_i\in A_u$ with respect to product $v$, the interactions between $a_i$ and all the aspects in $A_v$ are calculated and summed up:
\begin{equation}
\begin{array}{l}
\textbf{x}_{v,i} = \textbf{g}_v\odot\textbf{c}_i,\\
\textbf{g}_v = \sum\limits_{j \in {A_v}}{\textbf{c}_j}.
\end{array}
\end{equation}
As the interaction between two aspects represents their similarity, $\textbf{x}_{v,i}$ represents the overall similarity between the aspect $a_i$ and the product $v$. To measure the importance of different aspect $a_i\in A_u$, $\textbf{x}_{v,i}$ is used as aspect $a_i$'s input to the user-level attention layer. The attention layer is defined as:
\begin{equation}
\begin{array}{l}
{{\hat \alpha }_{u,v,i}} = {\textbf{w}_{att_2}}^T\textbf{x}_{v,i},\\
{\alpha_{u,v,i}} = \frac{{\rm{exp}}\left( {\hat \alpha}_{u,v,i} \right)} {{\sum\limits_{j \in {A_u}} {{\rm{exp}}\left( {{{\hat \alpha }_{u,v,j}}} \right)} } }.
\end{array}
\end{equation}
Here $\textbf{w}_{att_2}\in \mathbb{R}^{d_{a}}$ is a learnable vector, and ${\alpha }_{u,v,i}$ represents the importance of aspect $a_i\in A_u$ in user $u$'s preferences with regard to product $v$. This attention layer is different from the aspect-level attention layer defined in Equation (4) as $\textbf{w}_{att_1}$ and $\textbf{w}_{att_2}$ are two different vectors.

Finally, we compress the vector set $\{\textbf{h}_i\}_{a_i\in A_u}$ with a weighted sum pooling to generate a vector which represents user $u$'s overall satisfaction towards product $v$:
\begin{equation}
\textbf{y}_A(u,v) = \sum\limits_{i \in {A_u}} \alpha_{u,v,i}\textbf{h}_i.
\end{equation}
Here $\textbf{y}_A(u,v)\in \mathbb{R}^{d_a}$ is the output of Aspect Interactions Module.

\subsection{Global Interactions Part}
To model the implicit factors which are not mentioned in review text but have influence over users' satisfaction, AARM assigns a latent factor for every user and product respectively. In this module, embedding matrix $\textbf{W}^G_{U}\in \mathbb{R}^{d_{g}\times |U|}$ is defined to project user $u$ to $\textbf{p}_u$, and the embedding matrix $\textbf{W}^G_{V}\in \mathbb{R}^{d_{g}\times |V|}$ is defined to project product $v$ to $\textbf{q}_v$. These two embedding matrices are randomly initialized and tuned during the training for top-N recommendation. Then the global interaction between user $u$ and product $v$ is calculated in a way similar to that in vanilla latent factor models:
\begin{equation}
\textbf{y}_G(u,v)=\textbf{p}_u \odot \textbf{q}_v.
\end{equation}
Here $\textbf{y}_G(u,v)\in \mathbb{R}^{d_g}$ is the output of this part.

\subsection{Output Layer}
To merge information from the aforementioned two modules, ${\textbf{y}_{A}}(u,v)$ and $\textbf{y}_G(u,v)$ are concatenated into one vector. And a regression layer without an activation function is stacked above it:
\begin{equation}
\hat{y}(u,v)=\textbf{W}_{out}\left[ \begin{array}{l}{{\textbf{f}}_G(u,v)}\\{{\textbf{y}}_{A}}(u,v)\end{array} \right].
\end{equation}
Here $\textbf{W}_{out}$ belongs to $\mathbb{R}^{1\times (d_a+d_g)}$. $\hat{y}(u,v)$ represents user $u$'s overall satisfaction score towards product $v$.

\subsection{Learning}
In this paper, we binarize the ratings scores and train AARM with a learning-to-rank method. Ranking methods are widely used in information retrieval~\cite{Liu2009rankir,luo2018scalable,hong2017coherent} and recommendation models~\cite{rank2019tois,he2017neural}. In AARM, we use Bayesian Personalized Ranking (BPR) which is a pair-wise method. This makes AARM suitable for recommendation with implicit feedbacks. Given a user $u$, a triple ($u$, $v^+$, $v^-$) is constructed for pair-wise training. Here, $v^+$ refers to the product that $u$ has purchased, while $v^-$ refers to an unpurchased one. During training, the positive user-product pair ($u$, $v^+$) is drawn from rating set $R$, which is accompanied with one negative pair ($u$, $v^-$), where $v^-$ is randomly sampled from $u$'s unpurchased products. Intuitively, AARM should give higher satisfaction score to the positive pair ($u$, $v^+$) than the negative pair ($u$, $v^-$). Hence, the BPR optimization criterion is employed as the objective function of AARM:
\begin{equation}
{L_{bpr}} =  \frac{-1}{{|R|}}\sum\limits_{(u,{v^ + }) \in R} {\log (\sigma ({{\hat y}}(u,{v^ + }) - {{\hat y}}(u,{v^ - })))}.
\end{equation}
Here, $\sigma$ refers to the sigmoid function, and $|R|$ is the number of positive pairs ($u$,$v^+$) in $R$. 

To prevent the possible overfitting, $L^2$ regularization is used on user and product embedding matrix and the kernel matrix of the output layer. As shown in Equation (12), to implement the $L^2$ regularization, we first calculate the mean values of element-wise square of these three matrices. The results are then multiplied by the $L^2$ regularization coefficient $\lambda$ and added to the loss function:
\begin{equation}
L = L_{bpr} + \lambda *(\frac{||\textbf{W}^G_{U}||^2}{|\textbf{W}^G_{U}|} + \frac{||\textbf{W}^G_{V}||^2}{|\textbf{W}^G_{V}|} + \frac{||\textbf{W}_{out}||^2}{|\textbf{W}_{out}|}).
\end{equation}
Here $\lambda$ controls the $L^2$ regularization strength, $||\textbf{W}||$ refers to the $L^2$-norm of the matrix $\textbf{W}$, and $|\textbf{W}|$ refers to the number of elements in the matrix $\textbf{W}$. We minimize the loss function $L$ to fit AARM from data.

Besides $L^2$ regularization, we also use dropout \cite{srivastava2014dropout} to reduce overfitting. Dropout can prevent complex co-adaptations on training data by randomly dropping some units during training \cite{srivastava2014dropout}. Dropout is employed on the output of Global Interactions module and the output of Aspect Interactions module.

\textbf{Aspect Embedding Pre-training}. In our experiments, \emph{gensim}'s implementation\footnote{https://radimrehurek.com/gensim/} of Word2vec is used to train the aspect embeddings. Before training embeddings with Word2vec, we first construct a dictionary for every dataset and then segment the reviews of each dataset into lists of words or phrases according to this dictionary. All the aspects (in the form of words or phrases) of each dataset are added into the corresponding dictionary to make sure that the Word2vec tool can recognize all the aspects and train embedding vectors for them. For each dataset, all the reviews in the training set are used for the training of aspect embedding. These embedding vectors are used as the initial values of the aspect embedding matrix $\textbf{W}_A$, which would not be tuned during the training for top-N recommendation.
\section{Experiments}
\label{sec:experiments}
In this section, we design experiments to study the following research questions:
\begin{itemize}
\item \textbf{RQ1} Can AARM outperform state-of-the-art methods on top-N recommendation task?
\item \textbf{RQ2} Can the interactions between different aspects improve the performance of AARM?
\item \textbf{RQ3} Can the modeling of varied user interests improve the performance of AARM? 
\item \textbf{RQ4} How does the initialization and tuning strategy of aspect embedding influence the performance of AARM?
\item \textbf{RQ5} What are the contributions of the Global Interaction part and Aspect Interaction part in the overall performance of AARM?
\end{itemize}

In the rest of this section, we will first introduce experimental settings, and then successively answer the above research questions with not only quantitative experiments but also qualitative case studies.

\begin{table}[t]
	\begin{center}
		\caption{\textbf{Statistics of the experimental datasets.}}
		\label{tab:dataset}
        \scalebox{1.0}{
		\begin{tabular}{ |l |c |c |c |c |c| } \hline
			\textbf{Dataset} & \textbf{\#Rating} & \textbf{\#User} & \textbf{\#Product} & \textbf{Sparsity}\\ \hline 
			Movies and TV	&1,697,533 &123,960 & 50,052 & 0.0274\% \\ \hline
			CDs and Vinyl	&1,097,592 &75,258 &64,421 &0.0226\% \\ \hline
            Clothing, Shoes and Jewelry &278,677 &39,387 &23,033 &0.0307\% \\ \hline
            Cell Phones and Accessories &194,439 &27,879 &10,429 &0.0669\% \\ \hline
            Beauty &198,502 &22,363 &12,101 &0.0734\% \\ \hline
		\end{tabular}}
	\end{center}
\end{table}

\subsection{Datasets}
We use the "5-core" subsets from the publicly accessible ``Amazon product dataset''\footnote{http://jmcauley.ucsd.edu/data/amazon/} \cite{he2016ups} for experiments. Here the ``5-core'' means that each user and product in the subset has at least five reviews. Each record in the dataset is composed of five variables including \emph{user}, \emph{product}, \emph{rating}, \emph{textual review} and \emph{helpfulness votes}. In AARM, we only use \emph{user}, \emph{product} and \emph{textual review}. To follow the setting of baseline methods, in our pair-wise learning-to-rank framework, ratings are binarized to construct positive user-product pairs. We adopt five different product categories from the ``Amazon product dataset'', i.e., \textit{`Movies and TV'}, \textit{`CDs and Vinyl'}, \textit{`Clothing, Shoes and Jewelry'}, \textit{`Cell Phones and Accessories'} and \textit{`Beauty'}. Some detailed statistics including the sparsity and the number of ratings (\#Rating), users (\#User) and products (\#Product) of the five datasets are summarized in Table 1. Sparsity is defined as $\#Rating / (\#User\times \#Product)$. We can see that the five datasets are of different sizes and different levels of sparsity, which could cover different recommendation scenarios.

For each user, its 70\% records are randomly selected as training set, while the rest of 30\% records are put into test set. Particularly, we use the exact same splits and evaluation measures as the experimental settings in \cite{zhang2017joint}\footnote{We would like to thank the authors for sharing us with the datasets and specific splits.}. This is to guarantee that all the methods are evaluated on exactly the same settings for fair comparisons.

\subsection{Aspects from User Reviews}
Some detailed statistics of the aspects extracted from user reviews by \emph{Sentires} are shown in Table 2. We can see that the number of aspects (Aspect\#), the average number of aspects per user (Ave. \# Aspect/User) and the average number of aspects per product (Ave. \# Aspect/Product) in the five datasets are varied, which makes our experiments more comprehensive.
\begin{table}[t]
	\begin{center}
		\caption{\textbf{Statistics of aspects extracted from reviews.}}
		\label{tab:dataset}
		\begin{tabular}{ |l| c| c| c| }\hline
			\textbf{Dataset} & \textbf{\#Aspect} & \textbf{Ave. \#Aspect/User} & \textbf{Ave. \#Aspect//Product}\\ \hline
			Movies and TV	&2865 &14.72 &32.24  \\ \hline
			CDs and Vinyl	&4033 &31.04 &41.31  \\ \hline
            Clothing, Shoes and Jewelry &525 &7.04 &9.77 \\ \hline
            Cell Phones and Accessories &648 &6.93 &12.50 \\ \hline
            Beauty &691 &9.72 &13.13 \\ \hline
		\end{tabular}
	\end{center}
\end{table}

Table 3 shows some examples of the aspects extracted from each dataset. We did not conduct any post-processing on the extracted aspects. Although there are some noise words in the aspect collection, \emph{Sentires} is largely effective in extracting many meaningful aspects that correspond to important product features. And there are synonymous aspects like ``songwriters'' and ``composers'', and related aspects like ``smell'' and ``chocolate smell'', which would usually be treated as disparate product features in most existing aspect-level models.
\begin{table}[t]
	\begin{center}
		\caption{\textbf{Some examples of the automatically extracted aspects.}}
		\label{tab:dataset}
		\begin{tabular}{| l | l |} \hline
			\textbf{Dataset} & \textbf{Aspects}\\ \hline
			Movies and TV	&  3d movie, cast, halloween film, halloween movie, harden, \\ & melodrama, movie star, screen time, thrillers, zombie movie\\ \hline
			CDs and Vinyl	&  1980s, band, crooners, crooning, country musics, \\ & fingerwork, singers, rock fans, songwriters, composers \\ \hline
            Clothing, Shoes and Jewelry & color, cottony, diamonds, fit, price, \\ & presentation box, sleeve shirts, sleeve, traction, torso \\ \hline
            Cell Phones and Accessories & usb, accessory, a little, car chargers, car speaker, \\ & charge cycle, charge cycles, looks, plastic, quality \\ \hline
            Beauty & results, smell, chocolate smell, odor, ingredient, \\ &ingredients, face feeling, hair feeling, sheen, shampoos \\ \hline 
		\end{tabular}
	\end{center}
\end{table}

\subsection{Evaluation Protocols}
To generate a top-N recommendation list for user $u$, a model first estimates the scores of $u$'s candidate products, then ranks all the candidate products according to the scores and truncates the ranking list at $N$. In this paper, $u$'s candidate products include all the products in $u$'s test set and those that have not been purchased by $u$. In the evaluation, products in $u$'s test set would be used as ground truth. Following the settings in \cite{zhang2017joint}, we set $N=10$. Four standard metrics are used in the evaluation: Recall, Precision, Normalized Discounted Cumulative Gain (NDCG) and Hit Ratio (HT).

\textbf{Recall} is the percentage of products that has been recommended to the user in the products that has been purchased by the user:
\begin{equation}
Recall = \frac{n_{tp}}{n_{gt}},
\end{equation}
where $n_{tp}$ is the number of ground truth products in the recommendation list, and $n_{gt}$ is the number of ground truth products. We average the measure across all testing users.

\textbf{Precision} is the percentage of products which has been purchased by the user in the top-N recommendation list:
\begin{equation}
Precision = \frac{n_{tp}}{N}.
\end{equation}
The measure is averaged across all testing users.

\textbf{NDCG} is a measure when the positions of the purchased products in the recommendation list are considered. NDCG is based on the Discounted Cumulative Gain (DCG):
\begin{equation}
DCG = \sum \limits^{N}_{i=1}{\frac{2^{rel_i}-1}{log_2(i+1)}}.
\end{equation}
Here $rel_i$ is the graded relevance of the product at position $i$ of the recommendation list for a user. The NDCG of a user is then calculated as:
\begin{equation}
NDCG = \frac{DCG}{IDCG}.
\end{equation}
Here IDCG is the DCG of the ideal recommendation list where the user's ground truth products are all ranked at the top. We average NDCG across all testing users.

\textbf{HT} is defined as in the following equation where $n_{hit}$ is the number of users who has purchased at least one product in its recommendation list:
\begin{equation}
HT = \frac{n_{hit}}{|U|}.
\end{equation}

\subsection{Baselines and Parameter Settings}
We compare our method AARM with the following baselines.

\textbf{BPR-MF} \cite{Rendle:2009:BBP:1795114.1795167}. The matrix factorization (MF) based on Bayesian Personalized Ranking (BPR), which combines MF-model with a pair-wise learning to rank loss function, is a solid baseline for top-N recommendation. Only user-product interaction data is used in this method.

\textbf{BPR-HFT} \cite{mcauley2013hidden}. The Hidden Factor and Topics (HFT) model associates topics extracted from reviews with latent factors learned from numerical ratings. It is one of the state-of-the-art review-based recommendation methods. The original HFT model is a rating prediction method. BPR-HFT \cite{zhang2017joint} modifies HFT by adding a Bayesian Personalized Ranking loss on top of HFT to generate the top-N recommendation.

\textbf{GMF} \cite{he2017neural}. Generalized Matrix Factorization (GMF) is one of the state-of-the-art neural network based recommendation method which only utilizes user-product interaction records. In experiments, we directly use the released code by the authors \footnote{https://github.com/hexiangnan/neural\_collaborative\_filtering}.

\textbf{BPR-AFM} \cite{AFM2017}. Attentional Factorization Machine (AFM) is an improved variant of the famous factorization machine (FM) \cite{rendle2010fm}. Similar to our method, AFM uses a neural attention network to discriminate the importance of different feature interactions. The original version of AFM is designed for regression task and optimizes the squared loss. We modified AFM by adding a Bayesian Personalized Ranking loss on top of AFM to generate the top-N recommendation. Given a user and an item as input, we use the user identity, the item identity, the user's aspects and the item's aspects as features. Both the identity features and aspect features have corresponding embedding vectors in the model, which are randomly initialized and then fine-tuned during the training.

\textbf{DeepCoNN} \cite{Zheng:2017:JDM:3018661.3018665}. The Deep Cooperative Neural Network is one of the state-of-the-art deep learning methods for recommendation which utilizes reviews to build user and product representations. It uses the review-based user and product representations for rating prediction.

\textbf{JRL} \cite{zhang2017joint}. The Joint Representation Learning model is a state-of-the-art method which integrates different information sources with deep learning methods for top-N recommendation. Textual reviews, product images and numerical ratings are jointly used in JRL.

\textbf{JRL-Review} \cite{zhang2017joint}. JRL-Review is a single-view version of JRL which incorporates textual reviews for top-N recommendation. JRL-Review employs PV-DBOW model \cite{Le:2014:DRS:3044805.3045025} to learn the vector representations of users and products from their corresponding reviews. It is one of the state-of-the-art review-based recommendation methods.

\textbf{eJRL} \cite{zhang2017joint}. eJRL is another variant of JRL which jointly utilizes textual reviews, product images and numerical ratings for recommendation. The difference between them is that eJRL prevents information propagation among different information sources.

The hyper-parameters of baselines are tuned on training set with five-fold cross-validation. In particular, the dimension of latent factors (or embeddings) for BPR-MF, BPR-HFT and DeepCoNN is 100. For BPR-HFT, the number of topics is 10. For JRL, JRL-Review and eJRL, the embedding size is set as 300. For GMF and BPR-AFM, the size of all the embedding vectors is set as 128.

\textbf{Parameter Settings.}
We implemented our methods with Tensorflow\footnote{\url{https://www.tensorflow.org/}}. When padding user aspect set to the same size, the maximum size $M_u$ was defined as the 75\% quantile of the sizes of all user aspect sets. Similarly, the maximum size $M_v$ of product aspect set was defined as the 75\% quantile of the sizes of all product aspect sets. For embedding layers, we set the dimension $d_{g}$ of user and product embeddings in the global interactions module to 128; set the dimension $d_{a}$ of aspect embeddings to 128. AARM was optimized with mini-batch Adam \cite{kingma2014adam} because Adam uses adaptive learning rates for parameters with different update frequencies and converges faster than vanilla stochastic gradient descent. We tested the learning rate of [0.001, 0.003, 0.01]. For the coefficient of $L^2$ regularization, [0.0, 0.0001, 0.01, 0.1] was tested. To prevent overfitting, in dropout layers, the dropout rate was set to 0.5. When pre-training aspect embeddings with Word2Vec, the window size and the number of noise words for negative sampling are both 5.

The model was trained for a maximum of 300 epochs with early stopping. To build the validation set, 1000 users are randomly selected from the users in the training set. For each user, one of his purchased products is randomly drawn from training set as the ground truth product in validation set. And when evaluating the model on the validation set, for each user, all the products which are not paired with the user in training set are added to the candidate set. Then to build recommendation list for each user, products in the candidate set are ranked according to the estimated matching degrees between them and the user. The aforementioned four measures are used to evaluate the top-N recommendation lists and then averaged across all the validation users. For every 10 epoch, we will test the model's performance on the validation set. The training would be stopped if half of the four measures decreased for 40 successive epochs.

\begin{table*}[t]
	\small
	\begin{center}
		\caption{\textbf{The NDCG and hit ratio (HT) results of baselines and the proposed method for RQ1. Due to limitation of space, we present the name of dataset `Movies and TV' as `Movies', `CDs and Vinyl' as `CDs', `Clothing, Shoes and Jewelry' as `Clothings', `Cell Phones and Accessories' as `Cell Phones' for short. The best results are highlighted in bold. The improvements (or decrements for negative values) achieved by AARM compared with the best review-based baseline (Impr-JRL-Review) and the best multi-modal baseline (Impr-JRL or Impr-eJRL) are shown in the last 3 rows.}}
       \label{tab:performance1}
       \setlength{\tabcolsep}{2.5pt}
       \renewcommand{\arraystretch}{1}
        \scalebox{1.0}{
		\begin{tabular}{| l | c c | c c | c c | c c | c c |} \hline
			& \multicolumn{2}{c|}{\textbf{Movies}} & \multicolumn{2}{c|}{\textbf{CDs}} & \multicolumn{2}{c|}{\textbf{Clothings}} & \multicolumn{2}{c|}{\textbf{Cell Phones}} & \multicolumn{2}{c|}{\textbf{Beauty}} \\ \hline
            \textbf{Measures(\%)}&NDCG&HT &NDCG&HT &NDCG&HT &NDCG&HT &NDCG&HT \\ \hline
            BPR-MF &1.267&4.421 &2.009&8.554 &0.601&1.767 &1.998&5.273 &2.753&8.241 \\
            GMF &3.519	& 10.897	& 4.530	& 14.266	& 1.144	& 2.795	& 3.623	& 8.230	& 4.079	& 11.112\\ \hline
            BPR-HFT &2.092&6.378 &2.661&9.926 &1.067&2.872 &3.151&8.125 &2.934&8.268 \\
            DeepCoNN &3.800&10.522 &4.218&13.857 &1.310&3.286 &3.636&9.913 &3.359&9.807 \\
            BPR-AFM & 3.649	& 11.578 & 4.716	& 15.278	& 1.354	& 3.511	 &3.627	& 9.229	& 4.103	& 11.899\\
            JRL-Review &4.222&12.958 &5.286&16.592 &1.270&3.527 &4.184&10.632 &4.216&12.422 \\ \hline
            eJRL&4.405&13.292 &5.023&16.081 &1.523&4.182 &4.185&10.531 &3.896&11.090 \\
            JRL &4.334&13.245 &5.378&16.774 &1.735&4.634 &4.364&10.940 &4.396&12.776 \\ \hline
            \textbf{AARM} &\textbf{5.020} & \textbf{15.187} &\textbf{7.252} & \textbf{20.749} & \textbf{1.956} & \textbf{4.915} & \textbf{4.976} & \textbf{11.568} & \textbf{5.314} & \textbf{13.648} \\
            Impr-JRL-Review &18.901 	&17.202 	&37.193 	&25.054 	&54.094 	&39.354 	&18.929 	&8.804 	&26.044 	&9.870 \\
            Impr-eJRL &13.961 	&14.257 	&44.376 	&29.028 	&27.742 	&17.527 	&18.901 	&9.847 	&36.396 	&23.066 \\
            Impr-JRL &15.828 	&14.662 	&34.846 	&23.697 	&12.795 	&6.064 	&14.024 	&5.740 	&20.883 	&6.825\\ \hline
		\end{tabular}}
	\end{center}
\end{table*}

\begin{table*}[t]
        \begin{center}
        \caption{\textbf{The corresponding recall and precision results of baselines and the proposed method.}}
        \label{tab:performance2}
        \setlength{\tabcolsep}{1.0pt}
        \renewcommand{\arraystretch}{1}
        \scalebox{0.9}{
		\begin{tabular}{| l | c c | c c | c c | c c | c c |} \hline
			& \multicolumn{2}{c|}{\textbf{Movies}} & \multicolumn{2}{c|}{\textbf{CDs}} & \multicolumn{2}{c|}{\textbf{Clothings}} & \multicolumn{2}{c|}{\textbf{Cell Phones}} & \multicolumn{2}{c|}{\textbf{Beauty}} \\ \hline
            \textbf{Measures(\%)}&Recall&Precision &Recall&Precision &Recall&Precision &Recall&Precision &Recall&Precision \\ \hline
            BPR-MF &1.988&0.528 &2.679&1.085 &1.046&0.185 &3.258&0.595 &4.241&1.143 \\
            GMF & 5.169	& 1.306	& 6.280	& 1.844	& 1.832	& 0.299	& 5.751	& 0.931	& 6.291	& 1.439\\\hline
            BPR-HFT &3.255&0.776 &3.570&1.268 &1.819&0.297 &5.307&0.860 &4.459&1.132 \\ 
            DeepCoNN &4.671&0.886 &6.001&1.681 &2.332&0.229 &6.353&0.999 &5.429&1.200 \\
             BPR-AFM & 5.314	& 1.409	& 6.499	& 2.030	& 2.275	& 0.366	& 6.244	& 1.021	& 6.373	& 1.522\\
            JRL-Review &6.145&1.465 &7.454&2.079 &2.211&0.336 &7.275&1.062 &6.766&1.467 \\ \hline
            eJRL &6.289&1.521 &6.973&2.002 &2.679&0.396 &7.130&1.054 &6.010&1.355 \\
            JRL &6.334&1.492 &7.545&2.085 &2.989&0.442 &7.510&1.096 &6.949&1.546 \\ \hline
            \textbf{AARM} & \textbf{7.140} & \textbf{1.834} & \textbf{9.965} & \textbf{2.716} & \textbf{3.292} & \textbf{0.511} & \textbf{8.014} & \textbf{1.259} & \textbf{7.947} & \textbf{1.818}\\
            Impr-JRL-Review &16.192 	&25.188 	&33.687 	&30.640 	&48.892 	&52.083 	&10.158 	&18.550 	&17.455 	&24.777 \\
            Impr-eJRL &13.532 	&20.579 	&42.908 	&35.664 	&22.882 	&29.040 	&12.398 	&19.450 	&32.230 	&34.170 \\
            Impr-JRL &12.725 	&22.922 	&32.074 	&30.264 	&10.137 	&15.611 	&6.711 	&14.872 	&14.362 	&17.594 \\ \hline 
		\end{tabular}}
		\end{center}
\end{table*}

\subsection{Model Comparison (RQ1)}
Tables 4 and 5 show the performance of our method and baselines on top-N recommendation task.  The performances of rating-based methods (BPR-MF and GMF), review-based methods (BPR-HFT, DeepCoNN, BPR-AFM and JRL-Review), multi-modal methods (eJRL and JRL) and our method (AARM) are shown in the four blocks in each table from top to bottom. The last block of each table also presents the percentage of improvements (or decrements for negative values) achieved by AARM as compared with the best review-based baseline (Impr-JRL-Review) and the best multi-modal baseline (Impr-JRL or Impr-eJRL). The best results are highlighted in bold.  As we use the same split as \cite{zhang2017joint}, we directly reproduce their results of BPR-MF, BPR-HFT, DeepCoNN, JRL-Review, eJRL and JRL for fair comparisons. From Tables 4 and 5, we can see that:

(1) In general, neural network based methods outperform shallow models (e.g. BPR-MF and BPR-HFT). GMF, which only uses user-product interaction data, even largely outperforms BPR-HFT which incorporates reviews for recommendation. This might be attributed to the powerful representation learning capacity of neural models.

(2) Generally, review-based methods outperforms rating-based methods. All the review-based methods outperforms BPR-MF. Among neural network based methods, BPR-AFM and JRL-Review also outperforms GMF. This shows that review is an important information source to boost recommendation performance.

(3) Our proposed method AARM outperforms all the rating-based methods and review-based methods on all the datasets in terms of different metrics. Compared to these baselines, AARM make better use of the user-product interaction records and review texts. This is because of AARM's finer-grained modeling of aspect interactions, which simultaneously considers the interactions between different aspects and user's varied attentions towards aspects. In the following sections, we further analyze how the specific designs of AARM boost its recommendation performance.

(4) AARM also outperforms both of the multi-modal deep learning methods on all the datasets and on all the measures. It is surprising that our method outperforms these multi-modal deep learning methods which not only utilize review data but also leverage product image and numerical rating data for recommendation.
This further indicates that textual review is a very informative information source and AARM's finer-grained aspect modeling could effectively leveraged reviews for recommendation. In the following sections, we will discuss the contribution of each part of AARM by comparing AARM with its variants.

\subsection{Effect of Interactions between Different Aspects (RQ2)}
Previous aspect-based methods neglect the interactions between synonymous and similar aspects when making recommendations, and are limited by the sparsity of shared aspects in the reviews of users and products. AARM alleviates this problem by modeling the interactions between different aspects and using an attention module to capture the important aspect interactions. To verify the effect of this design, we compare AARM with its two variant, which are termed as ``A\_Inter'' and ``No-AspectAtt'' in Figure 3, under the same experimental settings.

\begin{figure}[t]
\centering
\includegraphics[width = 1.0\textwidth]{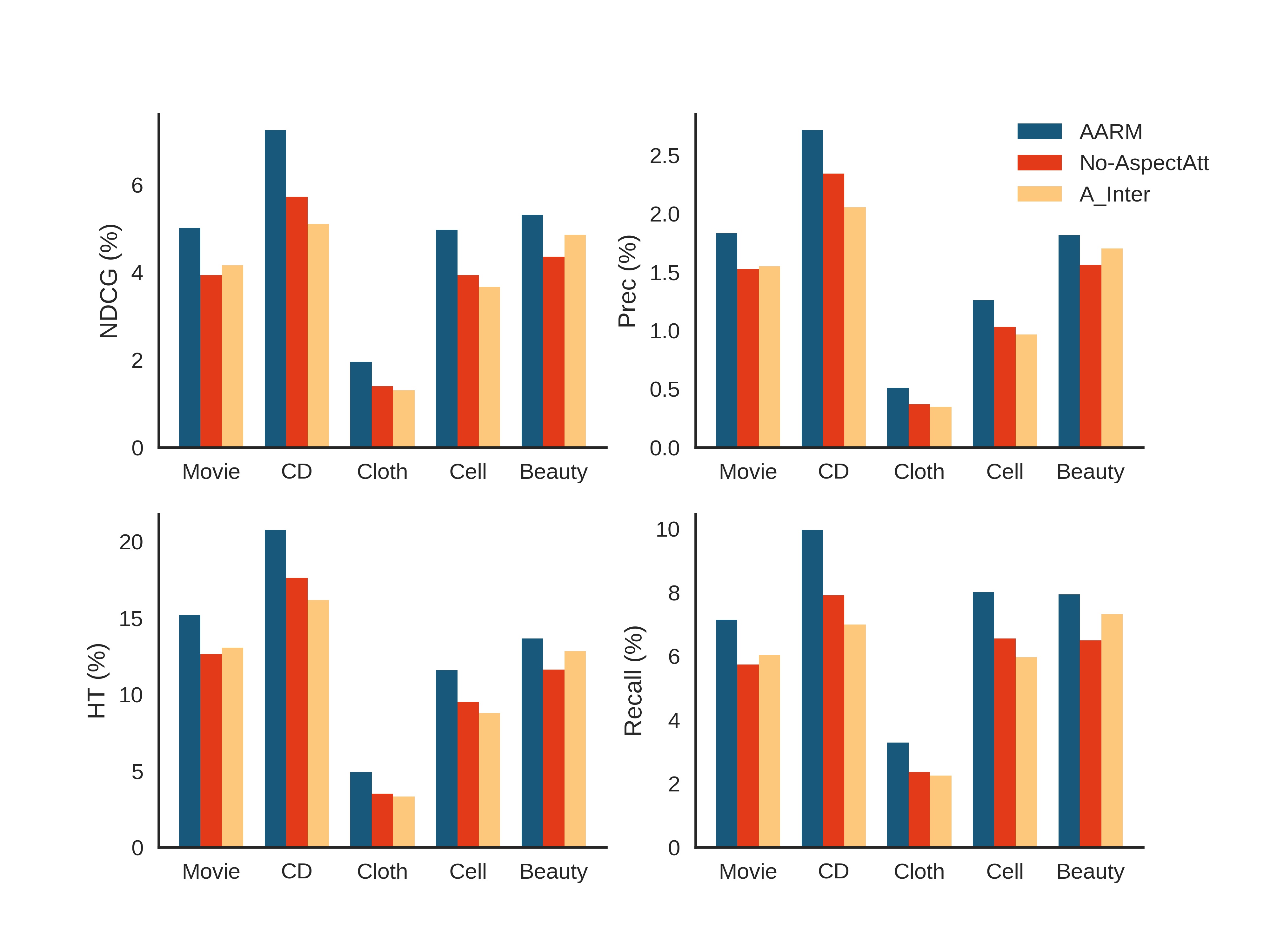}
\caption{\textbf{Performance of AARM, No-AspectAtt, and A\_Inter on five datasets for RQ2. Due to limitation of space, we present dataset `Movies and TV' as `Movie', `CDs and Vinyl' as `CD', `Clothing, Shoes and Jewelry' as `Cloth', `Cell Phones and Accessories' as `Cell' for short.}}
\end{figure}

As variants of AARM, the differences between AARM, No-AspectAtt and A\_Inter are in Aspect Interactions part. Given a user $u$ and a product $v$, A\_Inter only considers the interactions between shared aspects of $u$ and $v$, i.e., $a\in A_u\cap A_v$. Hence in the Aspect Interactions part of A\_Inter, we first calculate the intersection $A^{inter}_{u,v}$ of $A_u$ and $A_v$. To estimate $\textbf{h}_a$ which represents $u$'s preference to $v$ according to aspect $a\in A_u$, the Equations (3), (4) and (5) of AARM are replaced with the following equation:
\begin{equation}
\textbf{h}_a = \sum\limits_{i \in A^{inter}_{u,v}} {({\textbf{c}_i} \odot {\textbf{c}_i})(x_i)}.
\end{equation}
Here $x_i\in \{0,1\}$ is an indicator, where $x_i=0$ if $i$ is the meaningless aspect $<PAD>$ defined for padding. As A\_Inter only considers interactions between the same aspects, no aspect-level attention module is used here. In No-AspectAtt, the aspect-level attention layer are removed and the aspect interactions are directly summed up. The Equation (4) and (5) of AARM are replaced with the following equation:
\begin{equation}
\textbf{h}_i = \sum\limits_{j \in {A_v}} {({\textbf{c}_i}\odot {\textbf{c}_j})(x_ix_j)}.
\end{equation}

We evaluate A-Inter and No-AspectAtt's performance on top-N recommendation task and compare them with AARM in Figure 3. All the experimental settings are kept the same to ensure the reliability of results. As shown in Figure 3, AARM substantially outperforms A\_Inter and No-AspectAtt on all datasets in terms of all measures. Compared to A-Inter, the average improvements achieved by AARM are 39.401\% for NDCG, 37.427\% for recall, 32.823\% for HT and 33.593\% for precision. The results demonstrate the importance of modeling the interactions between different aspects and the effectiveness of our carefully designed aspect-level attentive layer. We will further perform qualitative analysis of the aspect-level attention layer in Section 4.10.

\subsection{Effect of Varied User Interest Modeling (RQ3)}
In the design of AARM, we assume that user's interests towards aspects are varied among different products. And an user-level attentive pooling layer (Equation (6), (7) and (8)), which simultaneously considers user, product and aspect information, is designed to capture user's different biases towards aspects when facing different products. To verify the effect of the user-level attention module, we design two variants of AARM, called A\_Static and No-UserAtt, and compare them with AARM on top-N recommendation task under the same settings. 

\begin{table}[]
\caption{\textbf{The NDCG and hit ratio (HT) results of AARM and its variants on five datasets for RQ3. We follow the short form convention adopted in Table 4 to name the datasets. The best performance of each measure on each dataset is highlighted in bold. The last block shows the percentage of improvements (or decrements for negative values) achieved by AARM compared with A\_static (Impr A\_static) and No-UserAtt (Impr No-UserAtt).}} 
\setlength{\tabcolsep}{2.0pt}
\begin{tabular}{|l|c|c|c|c|c|c|c|c|c|c|}
\hline
\multirow{2}{*}{\textbf{Measures(\%)}} & \multicolumn{2}{c|}{\textbf{Movies}} & \multicolumn{2}{c|}{\textbf{CDs}} & \multicolumn{2}{c|}{\textbf{Clothings}} & \multicolumn{2}{c|}{\textbf{Cell Phones}} & \multicolumn{2}{c|}{\textbf{Beauty}} \\ \cline{2-11}
                  & NDCG        & HT           & NDCG       & HT         & NDCG           & HT            & NDCG           & HT             & NDCG         & HT           \\ \hline
AARM & \textbf{5.020} & \textbf{15.187} & \textbf{7.252} & \textbf{20.749} & \textbf{1.957} & \textbf{4.915} & \textbf{4.976} & \textbf{11.568} & \textbf{5.314} & \textbf{13.648} \\ \hline
A\_Static &4.376 &13.318 &6.794 &19.567	 &1.898 &4.590 &4.728 &11.181 &4.918 &12.735 \\
No-UserAtt &4.290 &13.104 &6.700 &19.108 &1.310 &3.217 &4.685 &10.786 &4.739 &12.297 \\ \hline
Impr A\_static &14.717 	&14.034 	&6.741 	&6.041 	&3.109 	&7.081 	&5.245 	&3.461 	&8.052 	&7.169\\
Impr No-UserAtt &17.016 	&15.896 	&8.239 	&8.588 	&49.389 	&52.782 	&6.211 	&7.250 	&12.133 	&10.986\\
\hline
\end{tabular}
\end{table}

\begin{table}[]
\caption{\textbf{The corresponding precision and recall results of AARM and its variants on five datasets for RQ3.}} 
\setlength{\tabcolsep}{1.2pt}
\scalebox{0.95}{
\begin{tabular}{|l|c|c|c|c|c|c|c|c|c|c|}
\hline
\multirow{2}{*}{\textbf{Measures(\%)}} & \multicolumn{2}{c|}{\textbf{Movies}} & \multicolumn{2}{c|}{\textbf{CDs}} & \multicolumn{2}{c|}{\textbf{Clothings}} & \multicolumn{2}{c|}{\textbf{Cell Phones}} & \multicolumn{2}{c|}{\textbf{Beauty}} \\ \cline{2-11}
                  & Recall     & Precision     & Recall    & Precision   & Recall       & Precision       & Recall        & Precision       & Recall      & Precision     \\ \hline
AARM & \textbf{7.140} & \textbf{1.834} & \textbf{9.965} & \textbf{2.716} & \textbf{3.292} & \textbf{0.511} & \textbf{8.014} & \textbf{1.259} & \textbf{7.947} & \textbf{1.818} \\ \hline
A\_Static         &6.275	&1.588	&9.075	&2.470	&3.131	&0.476	&7.776 &1.219 &7.337	&1.699\\
No-UserAtt  & 6.076 &1.561 &8.953 &2.403 &2.193 &0.337 &7.583 &1.176 &7.046 &1.648 \\ \hline
Impr A\_static &13.785 	&15.491 	&9.807 	&9.960 	&5.142 	&7.353 	&3.061 	&3.281 	&8.314 	&7.004 \\ 
Impr No-UserAtt &17.512 	&17.489 	&11.303 	&13.025 	&50.114 	&51.632 	&5.684 	&7.058 	&12.787 	&10.316 \\

 \hline
\end{tabular}}
\end{table}

The differences between AARM, A\_Static and No-UserAtt are in the design of user-level attention module. A\_Static also assumes that user's interests towards different aspects are different. But different from AARM, A\_Static assumes that a user's interests towards aspects are fixed when facing different products. Therefore, the inputs of the user-level attention layer in A\_static do not consider the information of candidate products. When estimating user $u$'s interests towards its aspects, different from AARM, the input of the aspect $a_i\in A_u$ is designed as:
\begin{equation}
\begin{array}{l}
\textbf{x}_{u,i} = \textbf{g}_u\odot\textbf{c}_i,\\
\textbf{g}_u = \sum\limits_{j \in {A_u}}{\textbf{c}_j}.
\end{array}
\end{equation}
Here $\textbf{g}_u$ is the overall representation of aspects in $A_u$. And $\textbf{x}_{u,i}$, which represents a summation of the similarities between aspect $a_i$ and all the aspects in $A_u$, is aspect $a_i$'s input to the user-level attention layer.

Similar to Equation (7), The attention layer is defined as:
\begin{equation}
\begin{array}{l}
{{\hat \alpha }_{u,i}} = {\textbf{w}_{att_2}}^T\textbf{x}_{u,i},\\
{\alpha_{u,i}} = \frac{{\rm{exp}}\left( {\hat \alpha}_{u,i} \right)} {{\sum\limits_{j \in {A_u}} {{\rm{exp}}\left( {{{\hat \alpha }_{u,j}}} \right)} } }.
\end{array}
\end{equation}
Here $\textbf{w}_{att_2}\in \mathbb{R}^{d_{a}}$, and ${\alpha }_{u,i}$ represents the importance of aspect $a_i\in A_u$ with respect to the user $u$. From Equations (20) and (21), we can see that no product information is used in the user-level attention module.

Different from AARM, No-UserAtt assumes that a user would assign equal weights to its aspects when purchasing products. So instead of the user-level attentive pooling layer, No-UserAtt directly sums up the set of vectors $\{\textbf{h}_i\}_{a_i\in A_u}$ which represents the candidate product's performances on the aspects of user $u$:
\begin{equation}
\textbf{y}_A(u,v) = \sum\limits_{j \in {A_u}} \textbf{h}_j.
\end{equation}

As shown in Tables 6 and 7, AARM outperforms A\_static and No-UserAtt on all the datasets and on all the measures. Remind that the only differences between AARM and A\_static are the different assumptions about user attentions on aspects towards different products. From the results, we can see that AARM's varied user interests assumption is more reasonable as compared to the constant user interests assumption of A\_static. In real-life scenarios, a user could be interested in many different kinds of products and each product can be described by a specific set of aspects. Obviously the user will pay less attentions to the aspects which are not related to the current product. As no two products are exactly alike, a user's interests on the diverse aspects can be varied even for the products from the same category. We will further represent how the user-level attentive pooling works when facing different products in Section 4.10. 

In Tables 6 and 7, A\_static also outperforms No-UserAtt on all the datasets in general. As A\_static can be viewed as an enhanced version of No-UserAtt, where a fixed user interests model is added, we can see that identifying the different importance of aspects can boost the recommendation performance. This result is reasonable because different users have different tastes, and they would put different attentions to different product features.

\subsection{Effects of Initialization and Tuning Strategy of Aspect Embedding (RQ4)}
In AARM, the embeddings of aspects are first initialized with the vectors which are pre-trained with Word2vec on each dataset, and then transformed by the matrix $\textbf{W}_{trans}$. This is inspired by the findings in \cite{word2vec} that the word embeddings trained with Word2vec can retain the syntactic and semantic similarity relation between words. We keep the aspect embedding matrix $\textbf{W}_A$ fixed during the training of AARM for top-N recommendation while the matrix $\textbf{W}_{trans}$ are tunable during the training. We choose this tuning strategy because similar words will be shifted similarly as shown in \cite{goldberg2017neural}. 

There are also other two alternatives for the initialization and tuning strategies of aspect embedding matrix $\textbf{W}_A$. The first one is to randomly initialize the aspect embedding matrix and then tune it during the training for top-N recommendation. We conducted experiments under this setting and presented the results in Tables 8 and 9 in the row of ``Random+Tune''. The second choice is to initialize the aspect embedding matrix with pre-trained embeddings and then tune it during the training for top-N recommendation. The experiment results of the second settings is presented in Tables 8 and 9 in the row of ``Pretrain+Tune''.

\begin{table}[]
\caption{\textbf{The NDCG and hit ratio (HT) results of AARM and its variants on five datasets for RQ4.  We follow the short form convention adopted in Table 4 to name the datasets. The best performance of each measure on each dataset is highlighted in bold. The last block shows the percentage of improvements (or decrements for negative values) achieved by Random+Tune compared with Pretrain+Tune (Random vs. Pretrain).}} 
\setlength{\tabcolsep}{2.0pt}
\begin{tabular}{|l|c|c|c|c|c|c|c|c|c|c|}
\hline
\multirow{2}{*}{\textbf{Measures(\%)}} & \multicolumn{2}{c|}{\textbf{Movies}} & \multicolumn{2}{c|}{\textbf{CDs}} & \multicolumn{2}{c|}{\textbf{Clothings}} & \multicolumn{2}{c|}{\textbf{Cell Phones}} & \multicolumn{2}{c|}{\textbf{Beauty}} \\ \cline{2-11}
                  & NDCG        & HT           & NDCG       & HT         & NDCG           & HT            & NDCG           & HT             & NDCG         & HT           \\ \hline
AARM & \textbf{5.020} & \textbf{15.187} & \textbf{7.252} & \textbf{20.749} & \textbf{1.957} & \textbf{4.915} & \textbf{4.976} & \textbf{11.568} & \textbf{5.314} & \textbf{13.648} \\ \hline
Random+Tune & 4.607 &13.989 &6.709 &19.443 &1.487 &3.636 &4.354 &10.316 &4.794	&12.972 \\ 
Pretrain+Tune &4.764 &14.320 &6.744 &19.905 &0.802	&2.046 &4.210	&10.191 &4.658 &12.266 \\ \hline
Random vs. Pretrain &-3.296 	&-2.311 	&-0.519 	&-2.321 	&85.411 	&77.713 	&3.420 	&1.227 	&2.920 	&5.756 \\ \hline
\end{tabular}
\end{table}

\begin{table}[]
\caption{\textbf{The corresponding precision and recall results of AARM and its variants on five datasets for RQ4.}} 
\setlength{\tabcolsep}{1.2pt}
\scalebox{0.95}{
\begin{tabular}{|l|c|c|c|c|c|c|c|c|c|c|}
\hline
\multirow{2}{*}{\textbf{Measures(\%)}} & \multicolumn{2}{c|}{\textbf{Movies}} & \multicolumn{2}{c|}{\textbf{CDs}} & \multicolumn{2}{c|}{\textbf{Clothings}} & \multicolumn{2}{c|}{\textbf{Cell Phones}} & \multicolumn{2}{c|}{\textbf{Beauty}} \\ \cline{2-11}
                  & Recall     & Precision     & Recall    & Precision   & Recall       & Precision       & Recall        & Precision       & Recall      & Precision     \\ \hline
AARM & \textbf{7.140} & \textbf{1.834} & \textbf{9.965} & \textbf{2.716} & \textbf{3.292} & \textbf{0.511} & \textbf{8.014} & \textbf{1.259} & \textbf{7.947} & \textbf{1.818} \\ \hline
Random+Tune & 6.495 &1.667 &8.957 &2.428 &2.476 &0.382 &7.161 &1.135 &7.288 &1.706 \\ 
Pretrain+Tune &6.744 &1.719 &9.270 &2.616 &1.346 &0.216 &7.012 &1.110 &6.969 &1.647 \\ \hline
Random vs. Pretrain &-3.692 	&-3.025 	&-3.376 	&-7.187 	&83.952 	&76.852 	&2.125 	&2.252 	&4.577 	&3.582\\ \hline
\end{tabular}}
\end{table}

As shown in Tables 8 and 9, AARM with the ``pretraining + trainable linear transformation'' strategy outperforms Random+Tune and Pretrain+Tune on all the datasets and on all the measures. The results are reasonable because in the design of the attention layers in AARM, we assumed that the similarity between two aspects can be represented by the interaction between them. The capability of enabling similar words shifted similarly makes the ``pretraining + trainable linear transformation'' strategy more suitable for our task. 

Comparing the performance of Random+Tune with Pretrain+Tune in Tables 8 and 9, we can find that Pretrain+Tune outperforms Random+Tune in larger datasets like ``Movies and TV'' and ``CDs and Vinyl'' (refer to Table 1), while Random+Tune performs better in smaller datasets like ``Clothing, Shoes and Jewelry'', ``Cell Phones and Accessories'' and ``Beauty'' (refer to Table 1). This may be caused by the fact that when the training data is not sufficient, the Pretrain+Tune strategy may not be able to transform the pre-trained embeddings for the new task and thus lose the original similarity between words \cite{goldberg2017neural}. Random+Tune strategy which assigns a much smaller random initial values to embedding matrix is easier to be optimized for the new task in an end-to-end style. 

\subsection{Model Ablation: Effect of Global Module and Aspect Module (RQ5)}
In this section we examine the roles of the Global Interactions part and Aspect Interactions part in the results of AARM. As shown in Figure 1, given the user and product as input, the two parts of AARM worked separately. Then the outputs of these two parts are merged and input into the output layer to estimate the score. To verify the effect of the Aspect Interactions part, we remove the Global Interactions part from AARM, and directly input the result of Aspect Interactions part into the output layer. This variant of AARM is referred as ``Aspect Part'' in Tables 10 and 11. Similarly, another variant of AARM which is referred as ``Global Part'' in Tables 10 and 11 is constructed by removing the Aspect Interactions part from AARM to verify the effect of Global Interactions Part. 

\begin{table}[]
\caption{\textbf{The NDCG and hit ratio (HT) results of AARM and its variants on five datasets for RQ5. We follow the short form convention adopted in Table 4 to name the datasets. The best performance of each measure on each dataset is highlighted in bold. The last block shows the percentage of improvements (or decrements for negative values) achieved by Aspect Part compared with Global Part (Aspect vs. Global).}} 
\setlength{\tabcolsep}{2.0pt}
\begin{tabular}{|l|c|c|c|c|c|c|c|c|c|c|}
\hline
\multirow{2}{*}{\textbf{Measures(\%)}} & \multicolumn{2}{c|}{\textbf{Movies}} & \multicolumn{2}{c|}{\textbf{CDs}} & \multicolumn{2}{c|}{\textbf{Clothings}} & \multicolumn{2}{c|}{\textbf{Cell Phones}} & \multicolumn{2}{c|}{\textbf{Beauty}} \\ \cline{2-11}
                  & NDCG        & HT           & NDCG       & HT         & NDCG           & HT            & NDCG           & HT             & NDCG         & HT           \\ \hline
AARM & \textbf{5.020} & \textbf{15.187} & \textbf{7.252} & \textbf{20.749} & \textbf{1.957} & \textbf{4.915} & \textbf{4.976} & \textbf{11.568} & \textbf{5.314} & \textbf{13.648} \\ \hline
Global Part &3.035	&9.965	&4.860	&15.462 &1.084 &2.770 &3.492 &8.250 &4.199 &11.050 \\ 
Aspect Part & 2.401 &8.237 &5.200 &16.700 &1.677 &4.395 &3.006 &7.568 &3.781 &11.246\\ \hline
Aspect vs. Global &-20.890 	&-17.341 	&6.996 	&8.007 	&54.705 	&58.664 	&-13.918 	&-8.267 	&-9.955 	&1.774\\ \hline
\end{tabular}
\end{table}

\begin{table}[]
\caption{\textbf{The corresponding precision and recall results of AARM and its variants on five datasets for RQ5.}} 
\setlength{\tabcolsep}{1.2pt}
\scalebox{0.95}{
\begin{tabular}{|l|c|c|c|c|c|c|c|c|c|c|}
\hline
\multirow{2}{*}{\textbf{Measures(\%)}} & \multicolumn{2}{c|}{\textbf{Movies}} & \multicolumn{2}{c|}{\textbf{CDs}} & \multicolumn{2}{c|}{\textbf{Clothings}} & \multicolumn{2}{c|}{\textbf{Cell Phones}} & \multicolumn{2}{c|}{\textbf{Beauty}} \\ \cline{2-11}
                  & Recall     & Precision     & Recall    & Precision   & Recall       & Precision       & Recall        & Precision       & Recall      & Precision     \\ \hline
AARM & \textbf{7.140} & \textbf{1.834} & \textbf{9.965} & \textbf{2.716} & \textbf{3.292} & \textbf{0.511} & \textbf{8.014} & \textbf{1.259} & \textbf{7.947} & \textbf{1.818} \\ \hline
Global Part & 4.485 &1.206 &6.760 &2.057 &1.802 &0.295 &5.645 &0.895 &6.171 &1.507 \\ 
Aspect Part & 3.512 &0.936 &7.686 &2.020 &2.925 &0.451 &5.187 &0.794 &6.036 &1.249\\ \hline
Aspect vs. Global &-21.695 	&-22.388 	&13.698 	&-1.799 	&62.320 	&52.881 	&-8.113 	&-11.285 	&-2.188 	&-17.120 \\ \hline
\end{tabular}}
\end{table}

\begin{table}[]
\caption{\textbf{The distributions of the number of shared aspects between a user and a product on the five datasets. From left to right, the columns present the ratios of different user-product pairs which have specific numbers of shared aspects. Specially, the last column represents the ratio of user-product pairs which have more than five shared aspects.}}
\begin{tabular}{|l|c|c|c|c|c|c|c|}
\hline
{\textbf{Datasets}} & 0      & 1      & 2      & 3      & 4      & 5     & \textgreater{}5 \\ \hline
Cell Phones and Accessories & 26.34\% & 28.95\% & 19.73\% & 11.31\% & 6.09\%  & 3.26\% & 4.33\%           \\ \hline
Beauty    & 35.31\% & 29.85\% & 16.37\% & 8.21\%  & 4.22\%  & 2.29\% & 3.76\%           \\ \hline
Clothing, Shoes and Jewelry  & 12.09\% & 24.90\% & 25.50\% & 17.92\% & 10.07\% & 5.01\% & 4.50\%           \\ \hline
Movies and TV     & 30.98\% & 27.26\% & 15.54\% & 8.73\%  & 5.21\%  & 3.30\% & 8.98\%           \\ \hline
CDs and Vinyl        & 3.06\%  & 10.71\% & 13.91\% & 13.36\% & 11.39\% & 9.29\% & 38.27\%          \\ \hline
\end{tabular}
\end{table}

From Tables 10 and 11, we can find that AARM significantly outperforms Aspect Part and Global Part. This result indicates that our combination strategy based on concatenation is valid. And the Global Interactions part, which is designed to capture the user preferences that have not been mentioned in review texts, is an effective complement to the Aspect Interactions part.

As compared with Global Part, Aspect Part performs better in two datasets while falls behind in the other three datasets. Because Aspect Part connects users and products via the interactions between their aspects, its performance may be influenced by the number of interactions between related aspects. To verify this viewpoint, We traverse all the users and products in a dataset to construct all the possible user-product pairs, and then count the number of shared aspects of each user-product pair. A shared aspect of a user-product pair is a aspect which has been mentioned in both the user and the product's reviews. The distributions of the number of shared aspects of each user-product pair on the five datasets are shown in Table 12.

From Tables 10, 11 and 12, we can find that Aspect Part usually performs better on datasets which have more shared aspects between each user-product pair in general. For example, Aspect Part substantially outperforms Global Part in ``CDs and Vinyl'' and ``Clothing, Shoes and Jewelry'' datasets which have the smallest ratios of 0 shared aspects (see the 2nd column in the table). And for datasets ``Movies and TV'', ``Cell Phones and Accessories'' and ``Beauty'' where more than 20\% user-product pairs do not have any shared aspects, Global Part outperforms Aspect Part.

\subsection{Case Study of Attention Layers}
The user-level and aspect-level attention modules are important parts of AARM. The user-level attention module (refer to Equation 7) is employed to capture user's varied preferences on aspects. And the aspect-level attention module (refer to Equation 4) is designed to enhance the interactions between meaningful aspect pairs, like the interactions between the same or similar aspects, and reduce the influence of the interactions between the two irrelevant aspects. To illustrate the roles of these two attention modules in AARM, we randomly selected some examples for qualitative analysis. 

In Table 13, we show the user-level attention values of a user `A1P9UMP1XSE6MI' in ``Cell Phones and Accessories'' dataset when examining different products. The first column is the ids of four products in the dataset and their aspect sets. Each product has 15 aspects which is the 75\% quantile of the sizes of all product aspect sets in the dataset. The rest of columns show the aspects of the user (the second row from top to bottom) and the attention values that assigned to these aspects when facing aforementioned four products. From each product's aspect set, we can find that product `B00EOE6FUW' is a `usb charger', `B005HS5MKS' is a `bluetooth earpiece', and `B002VPE1NO' and `B00E8GYIRI' are the `shell case' of cell phones. The shared aspects of each user-product pair and corresponding attention values are highlighted in red. 

\begin{table*}[]
\centering
\caption{\textbf{A case study of the user-level attention module. The first column (from the left) shows ids and aspect sets of four products from the ``Cell Phones and Accessories'' dataset. The rest of columns show the aspects of the user (the second row from top to bottom) and the attention values assigned to these aspects when facing aforementioned four products. In each row, the aspects mentioned in both the user and product's reviews and their corresponding attention values are highlighted in red.}}
\scalebox{0.77}{
\setlength{\tabcolsep}{1.0pt}
\begin{tabular}{|l|c|c|c|c|c|c|c|c|c|}
\hline
\multicolumn{1}{|c|}{} & \multicolumn{9}{c|}{\textbf{Aspects of User A1P9UMP1XSE6MI}} \\ \hline
\multicolumn{1}{|c|}{\textbf{Products and Their Aspects}} & \multicolumn{1}{c|}{\begin{tabular}[c]{@{}c@{}}sound\\ quality\end{tabular}} & \multicolumn{1}{c|}{\begin{tabular}[c]{@{}c@{}}shell\\ case\end{tabular}} & \multicolumn{1}{c|}{grommets} & \multicolumn{1}{c|}{quality} & \multicolumn{1}{c|}{\begin{tabular}[c]{@{}c@{}}impact\\ protection\end{tabular}} & \multicolumn{1}{c|}{\begin{tabular}[c]{@{}c@{}}usb\\ cords\end{tabular}} & \multicolumn{1}{c|}{\begin{tabular}[c]{@{}c@{}}bluetooth\\ earpiece\end{tabular}} & \multicolumn{1}{c|}{\begin{tabular}[c]{@{}c@{}} usb\\ plug\end{tabular}} & \multicolumn{1}{c|}{grab} \\ \hline
\begin{tabular}[l]{@{}l@{}} B00EOE6FUW: \footnotesize{usb, usb cable, charging device, colors,} \\ \footnotesize{ cable, usb charger, car trip, {\color{red}{usb cords}}, usb end, nokia lumia,}\\ \footnotesize{usb chargers, car chargers, wiggle, ultra, {\color{red}{usb plug}}} \end{tabular} & 0.0013 & 0.0008 & 0.0058 & 0.0014 & 0.0003 & \color{red}\textbf{0.4406} & 0.0034 & \color{red}\textbf{0.5389} & 0.0075 \\ \hline 
\begin{tabular}[l]{@{}l@{}} B005HS5MKS: \footnotesize{peeve, {\color{red}{sound quality}}, sizes,} \\ \footnotesize{{\color{red}{bluetooth earpiece}}, downside, {\color{red}{quality}}, protection,} \footnotesize{looks} \end{tabular} & {\color{red}{0.4161}} & 0.0103 & 0.1416 & {\color{red}{0.1392}} & 0.0126 & 0.0371 & {\color{red}{0.1780}} & 0.0174 & 0.0477 \\ \hline
\begin{tabular}[l]{@{}l@{}} B002VPE1NO: \footnotesize{metallic, {\color{red}{shell case}}, shell, looks, grip,} \\ \footnotesize{finish, {\color{red}{impact protection}}, protection,iphone cases, } \\ \footnotesize{ {\color{red}{grommets}}, rubber strips, plastic, case w, armor,air case} \end{tabular} & 0.0109 & \color{red}\textbf{0.4785} & \color{red}\textbf{0.1309} & 0.0084 & \color{red}\textbf{0.1464} & 0.0199 & 0.0197 & 0.0102 & 0.1751 \\ \hline
\begin{tabular}[l]{@{}l@{}} B00E8GYIRI: \footnotesize{{\color{red}{impact protection}}, protection, shell, packing} \\ \footnotesize{snapon cases, plastic, plastic case, case, scuff, bulk, matte} \\ \footnotesize{phone protection, polycarbonate, iphone cases, {\color{red}{shell case}}} \end{tabular} & 0.0077 & \color{red}\textbf{0.6295}  & 0.0248 & 0.0042 & \color{red}\textbf{0.1929} & 0.0160 & 0.0144 & 0.0121 & 0.0984 \\ \hline
\end{tabular}}
\end{table*}

As shown in Table 13, when examining a product, the user-level attention module can find the aspects which are related to the product and assign higher attention values to them. First, all the shared aspects (highlighted in red) of each user-product pair are assigned much higher attention values. Second, the user-level attention module can assign higher values to aspects that are related to the product but have not been mentioned in the product's reviews. For example, when examining the shell cases `B002VPE1NO' and `B00E8GYIRI', `grab' is assigned higher weight although it is not in the product's aspect set. This is because that there are some related aspects of `grab' in the two products' aspect sets which are captured by our attention module (refer to Figure 4). 

The examples in Table 13 indicate why AARM can outperform A\_Static and No-UserAtt (refer to Tables 6 and 7). The user's aspect set consists of three unrelated kinds of aspects: 1) `sound quality', `quality' and `bluetooth earpiece'; 2) `usb cords' and `usb plug'; 3) `shell case', `grommets', `impact protection' and `grab'. In this case, No-UserAtt would assign same weights to aspect `bluetooth earpiece' and `shell case' when purchasing a bluetooth earpiece. And A\_Static would assign same weights to aspect `sound quality' no matter what kinds of products the user is purchasing. By identifying different aspects' different roles when purchasing different products, AARM achieved better performance.

\begin{figure*}[t]
\centering
\scalebox{1.0}{
\includegraphics[width = 1.0\textwidth]{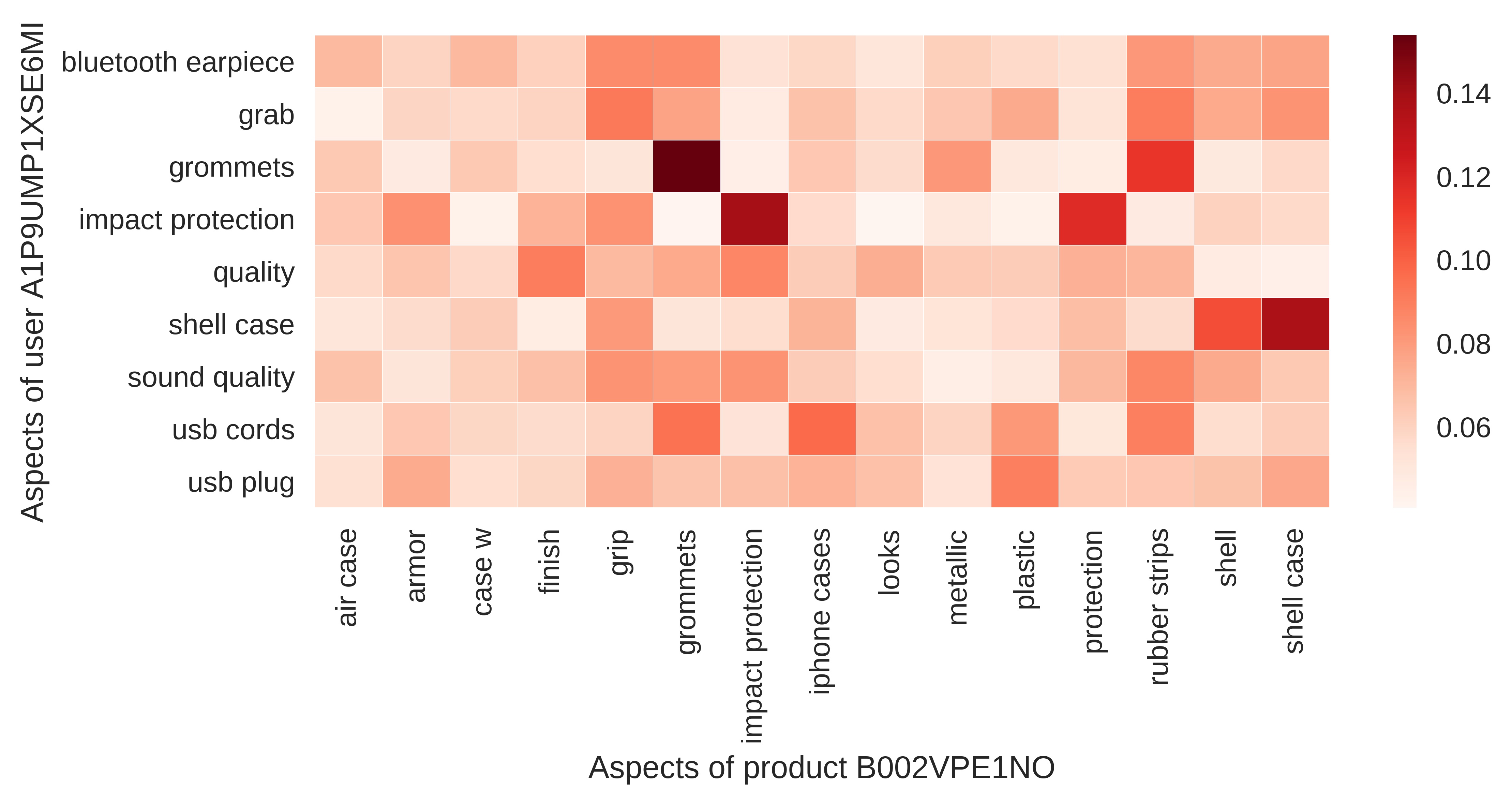}
}
\caption{\textbf{Heat map of aspect-level attention. The columns refer to aspects of the product while the rows refer to aspects of the user. Darker color in the grid cell means that higher attention value is assigned to the interaction between corresponding aspects by the aspect-level attention module.}}
\end{figure*}

Next we present how the aspect-level attention module finds the meaningful interactions (i.e., interactions between the shared aspects, synonymous aspects and similar aspects) from all the aspect interactions between a user and a product. In Figure 4, we show the aspect-level attention values of the interactions between aforementioned user `A1P9UMP1XSE6MI' and product `B002VPE1NO'. In the heat map, the columns refer to the product's aspects while the rows refer to the user's aspects. The color of each grid cell represents the attention value assigned to the corresponding interaction. The darker of the color in a grid cell, the higher of the attention value. 

First, we can see that interactions between the shared aspects like `grommets', `impact protection' and `shell case' are captured and assigned higher attention values. Second, the interactions between synonymous aspects are assigned higher weights as compared with unrelated ones. For example, (`shell case', `shell') is assigned the second highest attention value in the interactions between `shell case' and the product's aspects. Third, some interactions between similar aspects are captured. For example, in the interactions with `impact protection', the product's aspects `protection', `armor' and `grip' are assigned high attention values. Finally, for the user's aspects that are unrelated to the product (e.g. `usb plug'), their attention value distributions are more uniform compared to the shared and similar aspects. By assigning higher attention values to meaningful aspect interactions, AARM can alleviate the impact of noisy interactions and overcome the aspect sparsity problem.
\section{Conclusion and Future Work}
In this paper, we presented a Attentive Aspect-based Recommendation Model (AARM), which carefully capture the interactions between aspects extracted from reviews for recommendation. AARM first calculates the interactions between aspect embeddings to estimate how a product fits a user's requirements on each aspect, and then estimates the user's overall satisfactory on the product by synthesizing the product's performances on each aspect. To deal with the problem that the number of shared aspects between a user and a product is often limited, AARM takes the interactions between different aspects into consideration. With a well-designed aspect-level attention module, not only the shared aspects but also other related aspect pairs can be selected and assigned higher attention values. In addition, we hold the assumption that a user's interests towards aspects are varied when examining different products. To achieve the goal, an attention module which simultaneously considers user and product information is designed in AARM. In the experiments on five real-world datasets, AARM outperforms the state-of-the-art methods on the top-N recommendation task. In particular, compared with multi-modal (textual reviews, product images and numerical ratings) methods JRL and eJRL, AARM can still achieves better results in all datasets. To demonstrates the effectiveness of each component in AARM, a lot of quantitative experiments and qualitative case studies are conducted. 

In the future, we would like to extend our work in the following three ways: (1) Applying our method to capture the similarity relation between two different aspects to other recommendation scenes. By using the pre-trained aspect embedding, the aspect embedding transformation module and the aspect interaction layer, AARM can mimic the cosine similarity and capture the semantics and syntax similarities between two aspects. This strategy can also be used in other recommendation scenes (e.g. recommendation with tags or item metadata) to capture the relation between different elements (like tags or item categories). (2) Extracting aspects with neural network and combining it with AARM. In particular, we would like to jointly train the aspect extraction module and the recommendation module in an end-to-end style. Ideally, the end-to-end training could reduce noisy aspects and mine more domain-specific aspects. (3) Integrating aspect-level sentiment information in AARM. Aspect-level sentiment information is useful to identify user's likes and dislikes about product features. But existing methods usually use external tools for aspect-level sentiment analysis, which relies on the accuracy of these tools and is usually not able to deal with new reviews. We will study how to extract these sentiment information and integrate them into AARM with end-to-end learning.

\bibliographystyle{ACM-Reference-Format}
\bibliography{aarm}


\begin{thebibliography}{51}


\ifx \showCODEN    \undefined \def \showCODEN     #1{\unskip}     \fi
\ifx \showDOI      \undefined \def \showDOI       #1{#1}\fi
\ifx \showISBNx    \undefined \def \showISBNx     #1{\unskip}     \fi
\ifx \showISBNxiii \undefined \def \showISBNxiii  #1{\unskip}     \fi
\ifx \showISSN     \undefined \def \showISSN      #1{\unskip}     \fi
\ifx \showLCCN     \undefined \def \showLCCN      #1{\unskip}     \fi
\ifx \shownote     \undefined \def \shownote      #1{#1}          \fi
\ifx \showarticletitle \undefined \def \showarticletitle #1{#1}   \fi
\ifx \showURL      \undefined \def \showURL       {\relax}        \fi
\providecommand\bibfield[2]{#2}
\providecommand\bibinfo[2]{#2}
\providecommand\natexlab[1]{#1}
\providecommand\showeprint[2][]{arXiv:#2}

\bibitem[\protect\citeauthoryear{Bauman, Liu, and Tuzhilin}{Bauman
  et~al\mbox{.}}{2017}]%
        {Bauman:2017:ABR:3097983.3098170}
\bibfield{author}{\bibinfo{person}{Konstantin Bauman}, \bibinfo{person}{Bing
  Liu}, {and} \bibinfo{person}{Alexander Tuzhilin}.}
  \bibinfo{year}{2017}\natexlab{}.
\newblock \showarticletitle{Aspect based recommendations: Recommending items
  with the most valuable aspects based on user reviews}. In
  \bibinfo{booktitle}{\emph{Proceedings of the 23rd ACM SIGKDD International
  Conference on Knowledge Discovery and Data Mining}}. ACM,
  \bibinfo{pages}{717--725}.
\newblock


\bibitem[\protect\citeauthoryear{Cao, He, Miao, An, Yang, and Hong}{Cao
  et~al\mbox{.}}{2018}]%
        {CaoDaAGR}
\bibfield{author}{\bibinfo{person}{Da Cao}, \bibinfo{person}{Xiangnan He},
  \bibinfo{person}{Lianhai Miao}, \bibinfo{person}{Yahui An},
  \bibinfo{person}{Chao Yang}, {and} \bibinfo{person}{Richang Hong}.}
  \bibinfo{year}{2018}\natexlab{}.
\newblock \showarticletitle{Attentive group recommendation}. In
  \bibinfo{booktitle}{\emph{The 41st International ACM SIGIR Conference on
  Research and Development in Information Retrieval}}.
  \bibinfo{publisher}{ACM}, \bibinfo{address}{New York, NY, USA},
  \bibinfo{pages}{645--654}.
\newblock
\showISBNx{978-1-4503-5657-2}


\bibitem[\protect\citeauthoryear{Cao, He, Nie, Wei, Hu, Wu, and Chua}{Cao
  et~al\mbox{.}}{2017}]%
        {CaoDaTois}
\bibfield{author}{\bibinfo{person}{Da Cao}, \bibinfo{person}{Xiangnan He},
  \bibinfo{person}{Liqiang Nie}, \bibinfo{person}{Xiaochi Wei},
  \bibinfo{person}{Xia Hu}, \bibinfo{person}{Shunxiang Wu}, {and}
  \bibinfo{person}{Tat-Seng Chua}.} \bibinfo{year}{2017}\natexlab{}.
\newblock \showarticletitle{Cross-platform app recommendation by jointly
  modeling ratings and texts}.
\newblock \bibinfo{journal}{\emph{ACM Transactions on Information Systems}}
  \bibinfo{volume}{35}, \bibinfo{number}{4}, Article \bibinfo{articleno}{37}
  (\bibinfo{year}{2017}), \bibinfo{numpages}{37:1--37:27}~pages.
\newblock
\showISSN{1046-8188}


\bibitem[\protect\citeauthoryear{Catherine and Cohen}{Catherine and
  Cohen}{2017}]%
        {Catherine:2017:TLT:3109859.3109878}
\bibfield{author}{\bibinfo{person}{Rose Catherine} {and}
  \bibinfo{person}{William Cohen}.} \bibinfo{year}{2017}\natexlab{}.
\newblock \showarticletitle{Transnets: Learning to transform for
  recommendation}. In \bibinfo{booktitle}{\emph{Proceedings of the Eleventh ACM
  Conference on Recommender Systems}}. ACM, \bibinfo{pages}{288--296}.
\newblock


\bibitem[\protect\citeauthoryear{Chen, Zhang, Liu, and Ma}{Chen
  et~al\mbox{.}}{2018}]%
        {NARRE2018WWW}
\bibfield{author}{\bibinfo{person}{Chong Chen}, \bibinfo{person}{Min Zhang},
  \bibinfo{person}{Yiqun Liu}, {and} \bibinfo{person}{Shaoping Ma}.}
  \bibinfo{year}{2018}\natexlab{}.
\newblock \showarticletitle{Neural Attentional Rating Regression with
  Review-level Explanations}. In \bibinfo{booktitle}{\emph{Proceedings of the
  2018 World Wide Web Conference}}. \bibinfo{publisher}{International World
  Wide Web Conferences Steering Committee}, \bibinfo{address}{Republic and
  Canton of Geneva, Switzerland}, \bibinfo{pages}{1583--1592}.
\newblock


\bibitem[\protect\citeauthoryear{Chen, Zhang, Liu, and Ma}{Chen
  et~al\mbox{.}}{2019}]%
        {wsdm2019chen}
\bibfield{author}{\bibinfo{person}{Chong Chen}, \bibinfo{person}{Min Zhang},
  \bibinfo{person}{Yiqun Liu}, {and} \bibinfo{person}{Shaoping Ma}.}
  \bibinfo{year}{2019}\natexlab{}.
\newblock \showarticletitle{Social Attentional Memory Network: Modeling Aspect-
  and Friend-level Differences in Recommendation}. In
  \bibinfo{booktitle}{\emph{The eleventh ACM International Conference on Web
  Search and Data Mining}}.
\newblock


\bibitem[\protect\citeauthoryear{Chen, Zhang, He, Nie, Liu, and Chua}{Chen
  et~al\mbox{.}}{2017}]%
        {chen2017attentive}
\bibfield{author}{\bibinfo{person}{Jingyuan Chen}, \bibinfo{person}{Hanwang
  Zhang}, \bibinfo{person}{Xiangnan He}, \bibinfo{person}{Liqiang Nie},
  \bibinfo{person}{Wei Liu}, {and} \bibinfo{person}{Tat-Seng Chua}.}
  \bibinfo{year}{2017}\natexlab{}.
\newblock \showarticletitle{Attentive collaborative filtering: Multimedia
  recommendation with item-and component-level attention}. In
  \bibinfo{booktitle}{\emph{Proceedings of the 40th International ACM SIGIR
  Conference on Research and Development in Information Retrieval}}. ACM,
  \bibinfo{pages}{335--344}.
\newblock


\bibitem[\protect\citeauthoryear{Chen, Chen, and Wang}{Chen
  et~al\mbox{.}}{2015}]%
        {Chen2015}
\bibfield{author}{\bibinfo{person}{Li Chen}, \bibinfo{person}{Guanliang Chen},
  {and} \bibinfo{person}{Feng Wang}.} \bibinfo{year}{2015}\natexlab{}.
\newblock \showarticletitle{Recommender systems based on user reviews: the
  state of the art}.
\newblock \bibinfo{journal}{\emph{User Modeling and User-Adapted Interaction}}
  \bibinfo{volume}{25}, \bibinfo{number}{2} (\bibinfo{year}{2015}),
  \bibinfo{pages}{99--154}.
\newblock


\bibitem[\protect\citeauthoryear{Chen, Qin, Zhang, and Xu}{Chen
  et~al\mbox{.}}{2016}]%
        {Chen2016LRF}
\bibfield{author}{\bibinfo{person}{Xu Chen}, \bibinfo{person}{Zheng Qin},
  \bibinfo{person}{Yongfeng Zhang}, {and} \bibinfo{person}{Tao Xu}.}
  \bibinfo{year}{2016}\natexlab{}.
\newblock \showarticletitle{Learning to rank features for recommendation over
  multiple categories}. In \bibinfo{booktitle}{\emph{Proceedings of the 39th
  International ACM SIGIR Conference on Research and Development in Information
  Retrieval}}. ACM, \bibinfo{pages}{305--314}.
\newblock


\bibitem[\protect\citeauthoryear{Cheng, Chang, Zhu, Kanjirathinkal, and
  Kankanhalli}{Cheng et~al\mbox{.}}{2019}]%
        {cheng2019mmalfm}
\bibfield{author}{\bibinfo{person}{Zhiyong Cheng}, \bibinfo{person}{Xiaojun
  Chang}, \bibinfo{person}{Lei Zhu}, \bibinfo{person}{Rose~C Kanjirathinkal},
  {and} \bibinfo{person}{Mohan Kankanhalli}.} \bibinfo{year}{2019}\natexlab{}.
\newblock \showarticletitle{MMALFM: Explainable recommendation by leveraging
  reviews and images}.
\newblock \bibinfo{journal}{\emph{ACM Transactions on Information Systems
  (TOIS)}} \bibinfo{volume}{37}, \bibinfo{number}{2} (\bibinfo{year}{2019}),
  \bibinfo{pages}{16}.
\newblock


\bibitem[\protect\citeauthoryear{Cheng, Ding, He, Zhu, Song, and
  Kankanhalli}{Cheng et~al\mbox{.}}{2018a}]%
        {cheng2018ijcai}
\bibfield{author}{\bibinfo{person}{Zhiyong Cheng}, \bibinfo{person}{Ying Ding},
  \bibinfo{person}{Xiangnan He}, \bibinfo{person}{Lei Zhu},
  \bibinfo{person}{Xuemeng Song}, {and} \bibinfo{person}{Mohan Kankanhalli}.}
  \bibinfo{year}{2018}\natexlab{a}.
\newblock \showarticletitle{$A^3$NCF: An adaptive aspect attention model for
  rating prediction}. In \bibinfo{booktitle}{\emph{Proceedings of the
  Twenty-Seventh International Joint Conference on Artificial Intelligence}}.
  \bibinfo{publisher}{International Joint Conferences on Artificial
  Intelligence Organization}, \bibinfo{pages}{3748--3754}.
\newblock


\bibitem[\protect\citeauthoryear{Cheng, Ding, Zhu, and Kankanhalli}{Cheng
  et~al\mbox{.}}{2018b}]%
        {Cheng2018www}
\bibfield{author}{\bibinfo{person}{Zhiyong Cheng}, \bibinfo{person}{Ying Ding},
  \bibinfo{person}{Lei Zhu}, {and} \bibinfo{person}{Mohan Kankanhalli}.}
  \bibinfo{year}{2018}\natexlab{b}.
\newblock \showarticletitle{Aspect-aware latent factor model: Rating prediction
  with ratings and reviews}. In \bibinfo{booktitle}{\emph{Proceedings of the
  2018 World Wide Web Conference}}. International World Wide Web Conferences
  Steering Committee, \bibinfo{pages}{639--648}.
\newblock


\bibitem[\protect\citeauthoryear{Chin, Zhao, Joty, and Cong}{Chin
  et~al\mbox{.}}{2018}]%
        {chin2018anr}
\bibfield{author}{\bibinfo{person}{Jin~Yao Chin}, \bibinfo{person}{Kaiqi Zhao},
  \bibinfo{person}{Shafiq Joty}, {and} \bibinfo{person}{Gao Cong}.}
  \bibinfo{year}{2018}\natexlab{}.
\newblock \showarticletitle{ANR: Aspect-based Neural Recommender}. In
  \bibinfo{booktitle}{\emph{Proceedings of the 27th ACM International
  Conference on Information and Knowledge Management}}. ACM,
  \bibinfo{pages}{147--156}.
\newblock


\bibitem[\protect\citeauthoryear{Cremonesi, Koren, and Turrin}{Cremonesi
  et~al\mbox{.}}{2010}]%
        {Cremonesi:2010:PRA:1864708.1864721}
\bibfield{author}{\bibinfo{person}{Paolo Cremonesi}, \bibinfo{person}{Yehuda
  Koren}, {and} \bibinfo{person}{Roberto Turrin}.}
  \bibinfo{year}{2010}\natexlab{}.
\newblock \showarticletitle{Performance of recommender algorithms on top-n
  recommendation tasks}. In \bibinfo{booktitle}{\emph{Proceedings of the fourth
  ACM conference on Recommender systems}}. ACM, \bibinfo{pages}{39--46}.
\newblock


\bibitem[\protect\citeauthoryear{Dong and Smyth}{Dong and Smyth}{2017}]%
        {ijcai2017-674}
\bibfield{author}{\bibinfo{person}{Ruihai Dong} {and} \bibinfo{person}{Barry
  Smyth}.} \bibinfo{year}{2017}\natexlab{}.
\newblock \showarticletitle{User-based opinion-based recommendation}. In
  \bibinfo{booktitle}{\emph{Proceedings of the 26th International Joint
  Conference on Artificial Intelligence}}. \bibinfo{publisher}{AAAI Press},
  \bibinfo{pages}{4821--4825}.
\newblock


\bibitem[\protect\citeauthoryear{Ebesu, Shen, and Fang}{Ebesu
  et~al\mbox{.}}{2018}]%
        {Ebesu2018CMN}
\bibfield{author}{\bibinfo{person}{Travis Ebesu}, \bibinfo{person}{Bin Shen},
  {and} \bibinfo{person}{Yi Fang}.} \bibinfo{year}{2018}\natexlab{}.
\newblock \showarticletitle{Collaborative Memory Network for Recommendation
  Systems}. In \bibinfo{booktitle}{\emph{The 41st International ACM SIGIR
  Conference on Research and Development in Information Retrieval}}
  \emph{(\bibinfo{series}{SIGIR '18})}. \bibinfo{publisher}{ACM},
  \bibinfo{address}{New York, NY, USA}, \bibinfo{pages}{515--524}.
\newblock


\bibitem[\protect\citeauthoryear{Ganu, Kakodkar, and Marian}{Ganu
  et~al\mbox{.}}{2013}]%
        {Ganu:2013:IQP:2379838.2379890}
\bibfield{author}{\bibinfo{person}{Gayatree Ganu}, \bibinfo{person}{Yogesh
  Kakodkar}, {and} \bibinfo{person}{Am{\'e}Lie Marian}.}
  \bibinfo{year}{2013}\natexlab{}.
\newblock \showarticletitle{Improving the quality of predictions using textual
  information in online user reviews}.
\newblock \bibinfo{journal}{\emph{Information Systems}} \bibinfo{volume}{38},
  \bibinfo{number}{1} (\bibinfo{year}{2013}), \bibinfo{pages}{1--15}.
\newblock


\bibitem[\protect\citeauthoryear{Goldberg}{Goldberg}{2017}]%
        {goldberg2017neural}
\bibfield{author}{\bibinfo{person}{Yoav Goldberg}.}
  \bibinfo{year}{2017}\natexlab{}.
\newblock \showarticletitle{Neural network methods for natural language
  processing}.
\newblock \bibinfo{journal}{\emph{Synthesis Lectures on Human Language
  Technologies}} \bibinfo{volume}{10}, \bibinfo{number}{1}
  (\bibinfo{year}{2017}), \bibinfo{pages}{1--309}.
\newblock


\bibitem[\protect\citeauthoryear{Guo, Cheng, Nie, Xu, and Kankanhalli}{Guo
  et~al\mbox{.}}{2018}]%
        {guo2018mm}
\bibfield{author}{\bibinfo{person}{Yangyang Guo}, \bibinfo{person}{Zhiyong
  Cheng}, \bibinfo{person}{Liqiang Nie}, \bibinfo{person}{Xin-Shun Xu}, {and}
  \bibinfo{person}{Mohan Kankanhalli}.} \bibinfo{year}{2018}\natexlab{}.
\newblock \showarticletitle{Multi-modal preference modeling for product
  search}. In \bibinfo{booktitle}{\emph{Proceedings of the 2018 ACM on
  Multimedia Conference}}. ACM.
\newblock


\bibitem[\protect\citeauthoryear{He and McAuley}{He and McAuley}{2016}]%
        {he2016ups}
\bibfield{author}{\bibinfo{person}{Ruining He} {and} \bibinfo{person}{Julian
  McAuley}.} \bibinfo{year}{2016}\natexlab{}.
\newblock \showarticletitle{Ups and downs: Modeling the visual evolution of
  fashion trends with one-class collaborative filtering}. In
  \bibinfo{booktitle}{\emph{proceedings of the 25th international conference on
  world wide web}}. International World Wide Web Conferences Steering
  Committee, \bibinfo{pages}{507--517}.
\newblock


\bibitem[\protect\citeauthoryear{He, Chen, Kan, and Chen}{He
  et~al\mbox{.}}{2015}]%
        {He:2015:TRE:2806416.2806504}
\bibfield{author}{\bibinfo{person}{Xiangnan He}, \bibinfo{person}{Tao Chen},
  \bibinfo{person}{Min-Yen Kan}, {and} \bibinfo{person}{Xiao Chen}.}
  \bibinfo{year}{2015}\natexlab{}.
\newblock \showarticletitle{TriRank: Review-aware explainable recommendation by
  modeling aspects}. In \bibinfo{booktitle}{\emph{Proceedings of the 24th ACM
  International on Conference on Information and Knowledge Management}}. ACM,
  \bibinfo{pages}{1661--1670}.
\newblock


\bibitem[\protect\citeauthoryear{He and Chua}{He and Chua}{2017}]%
        {NFM2017}
\bibfield{author}{\bibinfo{person}{Xiangnan He} {and} \bibinfo{person}{Tat-Seng
  Chua}.} \bibinfo{year}{2017}\natexlab{}.
\newblock \showarticletitle{Neural factorization machines for sparse predictive
  analytics}. In \bibinfo{booktitle}{\emph{Proceedings of the 40th
  International ACM SIGIR Conference on Research and Development in Information
  Retrieval}}. \bibinfo{publisher}{ACM}, \bibinfo{pages}{355--364}.
\newblock


\bibitem[\protect\citeauthoryear{He, He, Song, Liu, Jiang, and Chua}{He
  et~al\mbox{.}}{2018}]%
        {he2018nais}
\bibfield{author}{\bibinfo{person}{Xiangnan He}, \bibinfo{person}{Zhankui He},
  \bibinfo{person}{Jingkuan Song}, \bibinfo{person}{Zhenguang Liu},
  \bibinfo{person}{Yu-Gang Jiang}, {and} \bibinfo{person}{Tat-Seng Chua}.}
  \bibinfo{year}{2018}\natexlab{}.
\newblock \showarticletitle{Nais: Neural attentive item similarity model for
  recommendation}.
\newblock \bibinfo{journal}{\emph{IEEE Transactions on Knowledge and Data
  Engineering}} \bibinfo{volume}{30}, \bibinfo{number}{12}
  (\bibinfo{year}{2018}), \bibinfo{pages}{2354--2366}.
\newblock


\bibitem[\protect\citeauthoryear{He, Liao, Zhang, Nie, Hu, and Chua}{He
  et~al\mbox{.}}{2017}]%
        {he2017neural}
\bibfield{author}{\bibinfo{person}{Xiangnan He}, \bibinfo{person}{Lizi Liao},
  \bibinfo{person}{Hanwang Zhang}, \bibinfo{person}{Liqiang Nie},
  \bibinfo{person}{Xia Hu}, {and} \bibinfo{person}{Tat-Seng Chua}.}
  \bibinfo{year}{2017}\natexlab{}.
\newblock \showarticletitle{Neural collaborative filtering}. In
  \bibinfo{booktitle}{\emph{Proceedings of the 26th International Conference on
  World Wide Web}}. International World Wide Web Conferences Steering
  Committee, \bibinfo{pages}{173--182}.
\newblock


\bibitem[\protect\citeauthoryear{Hong, Li, Cai, Tao, Wang, and Tian}{Hong
  et~al\mbox{.}}{2017}]%
        {hong2017coherent}
\bibfield{author}{\bibinfo{person}{Richang Hong}, \bibinfo{person}{Lei Li},
  \bibinfo{person}{Junjie Cai}, \bibinfo{person}{Dapeng Tao},
  \bibinfo{person}{Meng Wang}, {and} \bibinfo{person}{Qi Tian}.}
  \bibinfo{year}{2017}\natexlab{}.
\newblock \showarticletitle{Coherent semantic-visual indexing for large-scale
  image retrieval in the cloud}.
\newblock \bibinfo{journal}{\emph{IEEE Transactions on Image Processing}}
  \bibinfo{volume}{26}, \bibinfo{number}{9} (\bibinfo{year}{2017}),
  \bibinfo{pages}{4128--4138}.
\newblock


\bibitem[\protect\citeauthoryear{Kingma and Ba}{Kingma and Ba}{2014}]%
        {kingma2014adam}
\bibfield{author}{\bibinfo{person}{Diederik~P Kingma} {and}
  \bibinfo{person}{Jimmy Ba}.} \bibinfo{year}{2014}\natexlab{}.
\newblock \showarticletitle{Adam: A method for stochastic optimization}.
\newblock \bibinfo{journal}{\emph{arXiv preprint arXiv:1412.6980}}
  (\bibinfo{year}{2014}).
\newblock


\bibitem[\protect\citeauthoryear{Koren, Bell, and Volinsky}{Koren
  et~al\mbox{.}}{2009}]%
        {5197422}
\bibfield{author}{\bibinfo{person}{Y. Koren}, \bibinfo{person}{R. Bell}, {and}
  \bibinfo{person}{C. Volinsky}.} \bibinfo{year}{2009}\natexlab{}.
\newblock \showarticletitle{Matrix factorization techniques for recommender
  systems}.
\newblock \bibinfo{journal}{\emph{Computer}} \bibinfo{volume}{42},
  \bibinfo{number}{8} (\bibinfo{year}{2009}), \bibinfo{pages}{30--37}.
\newblock


\bibitem[\protect\citeauthoryear{Le and Mikolov}{Le and Mikolov}{2014}]%
        {Le:2014:DRS:3044805.3045025}
\bibfield{author}{\bibinfo{person}{Quoc Le} {and} \bibinfo{person}{Tomas
  Mikolov}.} \bibinfo{year}{2014}\natexlab{}.
\newblock \showarticletitle{Distributed representations of sentences and
  documents}. In \bibinfo{booktitle}{\emph{International Conference on Machine
  Learning}}. \bibinfo{pages}{1188--1196}.
\newblock


\bibitem[\protect\citeauthoryear{Liu}{Liu}{2009}]%
        {Liu2009rankir}
\bibfield{author}{\bibinfo{person}{Tie-Yan Liu}.}
  \bibinfo{year}{2009}\natexlab{}.
\newblock \showarticletitle{Learning to Rank for Information Retrieval}.
\newblock \bibinfo{journal}{\emph{Found. Trends Inf. Retr.}}
  \bibinfo{volume}{3}, \bibinfo{number}{3} (\bibinfo{date}{March}
  \bibinfo{year}{2009}), \bibinfo{pages}{225--331}.
\newblock


\bibitem[\protect\citeauthoryear{Luo, Wu, and Xu}{Luo et~al\mbox{.}}{2018}]%
        {luo2018scalable}
\bibfield{author}{\bibinfo{person}{Xin Luo}, \bibinfo{person}{Ye Wu}, {and}
  \bibinfo{person}{Xin-Shun Xu}.} \bibinfo{year}{2018}\natexlab{}.
\newblock \showarticletitle{Scalable supervised discrete hashing for
  large-scale search}. In \bibinfo{booktitle}{\emph{Proceedings of the 2018
  World Wide Web Conference on World Wide Web}}. International World Wide Web
  Conferences Steering Committee, \bibinfo{pages}{1603--1612}.
\newblock


\bibitem[\protect\citeauthoryear{McAuley and Leskovec}{McAuley and
  Leskovec}{2013}]%
        {mcauley2013hidden}
\bibfield{author}{\bibinfo{person}{Julian McAuley} {and} \bibinfo{person}{Jure
  Leskovec}.} \bibinfo{year}{2013}\natexlab{}.
\newblock \showarticletitle{Hidden factors and hidden topics: understanding
  rating dimensions with review text}. In \bibinfo{booktitle}{\emph{Proceedings
  of the 7th ACM conference on Recommender systems}}. ACM,
  \bibinfo{pages}{165--172}.
\newblock


\bibitem[\protect\citeauthoryear{Meng, Wang, Liu, and Zhang}{Meng
  et~al\mbox{.}}{2018}]%
        {MIRROR2018AAAI}
\bibfield{author}{\bibinfo{person}{Xuying Meng}, \bibinfo{person}{Suhang Wang},
  \bibinfo{person}{Huan Liu}, {and} \bibinfo{person}{Yujun Zhang}.}
  \bibinfo{year}{2018}\natexlab{}.
\newblock \showarticletitle{Exploiting Emotion on Reviews for Recommender
  Systems}. In \bibinfo{booktitle}{\emph{AAAI Conference on Artificial
  Intelligence}}.
\newblock


\bibitem[\protect\citeauthoryear{Mikolov, Chen, Corrado, and Dean}{Mikolov
  et~al\mbox{.}}{2013}]%
        {word2vec}
\bibfield{author}{\bibinfo{person}{Tomas Mikolov}, \bibinfo{person}{Kai Chen},
  \bibinfo{person}{Greg Corrado}, {and} \bibinfo{person}{Jeffrey Dean}.}
  \bibinfo{year}{2013}\natexlab{}.
\newblock \showarticletitle{Efficient estimation of word representations in
  vector space}.
\newblock \bibinfo{journal}{\emph{arXiv preprint arXiv:1301.3781}}
  (\bibinfo{year}{2013}).
\newblock


\bibitem[\protect\citeauthoryear{Pan, Yang, Cai, Chen, Zhang, Peng, and
  Ming}{Pan et~al\mbox{.}}{2019}]%
        {rank2019tois}
\bibfield{author}{\bibinfo{person}{Weike Pan}, \bibinfo{person}{Qiang Yang},
  \bibinfo{person}{Wanling Cai}, \bibinfo{person}{Yaofeng Chen},
  \bibinfo{person}{Qing Zhang}, \bibinfo{person}{Xiaogang Peng}, {and}
  \bibinfo{person}{Zhong Ming}.} \bibinfo{year}{2019}\natexlab{}.
\newblock \showarticletitle{Transfer to Rank for Heterogeneous One-Class
  Collaborative Filtering}.
\newblock \bibinfo{journal}{\emph{ACM Trans. Inf. Syst.}} \bibinfo{volume}{37},
  \bibinfo{number}{1}, Article \bibinfo{articleno}{10} (\bibinfo{date}{Jan.}
  \bibinfo{year}{2019}), \bibinfo{numpages}{20}~pages.
\newblock


\bibitem[\protect\citeauthoryear{Pero and Horv{\'a}th}{Pero and
  Horv{\'a}th}{2013}]%
        {opinion_driven_mf}
\bibfield{author}{\bibinfo{person}{{\v{S}}tefan Pero} {and}
  \bibinfo{person}{Tom{\'a}{\v{s}} Horv{\'a}th}.}
  \bibinfo{year}{2013}\natexlab{}.
\newblock \showarticletitle{Opinion-driven matrix factorization for rating
  prediction}. In \bibinfo{booktitle}{\emph{International Conference on User
  Modeling, Adaptation, and Personalization}}. Springer,
  \bibinfo{pages}{1--13}.
\newblock


\bibitem[\protect\citeauthoryear{Rendle}{Rendle}{2010}]%
        {rendle2010fm}
\bibfield{author}{\bibinfo{person}{Steffen Rendle}.}
  \bibinfo{year}{2010}\natexlab{}.
\newblock \showarticletitle{Factorization machines}. In
  \bibinfo{booktitle}{\emph{Proceedings of the 2010 IEEE International
  Conference on Data Mining}}. \bibinfo{publisher}{IEEE Computer Society},
  \bibinfo{pages}{995--1000}.
\newblock


\bibitem[\protect\citeauthoryear{Rendle, Freudenthaler, Gantner, and
  Schmidt-Thieme}{Rendle et~al\mbox{.}}{2009}]%
        {Rendle:2009:BBP:1795114.1795167}
\bibfield{author}{\bibinfo{person}{Steffen Rendle}, \bibinfo{person}{Christoph
  Freudenthaler}, \bibinfo{person}{Zeno Gantner}, {and} \bibinfo{person}{Lars
  Schmidt-Thieme}.} \bibinfo{year}{2009}\natexlab{}.
\newblock \showarticletitle{BPR: bayesian personalized ranking from implicit
  feedback}. In \bibinfo{booktitle}{\emph{Proceedings of the twenty-fifth
  conference on uncertainty in artificial intelligence}}. AUAI Press,
  \bibinfo{pages}{452--461}.
\newblock


\bibitem[\protect\citeauthoryear{Salton, Wong, and Yang}{Salton
  et~al\mbox{.}}{1975}]%
        {salton1975vector}
\bibfield{author}{\bibinfo{person}{Gerard Salton}, \bibinfo{person}{Anita
  Wong}, {and} \bibinfo{person}{Chung-Shu Yang}.}
  \bibinfo{year}{1975}\natexlab{}.
\newblock \showarticletitle{A vector space model for automatic indexing}.
\newblock \bibinfo{journal}{\emph{Commun. ACM}} \bibinfo{volume}{18},
  \bibinfo{number}{11} (\bibinfo{year}{1975}), \bibinfo{pages}{613--620}.
\newblock


\bibitem[\protect\citeauthoryear{Srivastava, Hinton, Krizhevsky, Sutskever, and
  Salakhutdinov}{Srivastava et~al\mbox{.}}{2014}]%
        {srivastava2014dropout}
\bibfield{author}{\bibinfo{person}{Nitish Srivastava},
  \bibinfo{person}{Geoffrey~E Hinton}, \bibinfo{person}{Alex Krizhevsky},
  \bibinfo{person}{Ilya Sutskever}, {and} \bibinfo{person}{Ruslan
  Salakhutdinov}.} \bibinfo{year}{2014}\natexlab{}.
\newblock \showarticletitle{Dropout: a simple way to prevent neural networks
  from overfitting.}
\newblock \bibinfo{journal}{\emph{The Journal of Machine Learning Research}}
  \bibinfo{volume}{15}, \bibinfo{number}{1} (\bibinfo{year}{2014}),
  \bibinfo{pages}{1929--1958}.
\newblock


\bibitem[\protect\citeauthoryear{Tan, Wan, Liu, and Xiao}{Tan
  et~al\mbox{.}}{2018}]%
        {Tois2018Quote}
\bibfield{author}{\bibinfo{person}{Jiwei Tan}, \bibinfo{person}{Xiaojun Wan},
  \bibinfo{person}{Hui Liu}, {and} \bibinfo{person}{Jianguo Xiao}.}
  \bibinfo{year}{2018}\natexlab{}.
\newblock \showarticletitle{QuoteRec: Toward quote recommendation for writing}.
\newblock \bibinfo{journal}{\emph{ACM Transactions on Information Systems}}
  \bibinfo{volume}{36}, \bibinfo{number}{3} (\bibinfo{year}{2018}),
  \bibinfo{pages}{34:1--34:36}.
\newblock


\bibitem[\protect\citeauthoryear{Tan, Zhang, Liu, and Ma}{Tan
  et~al\mbox{.}}{2016}]%
        {rblt2016ijcai}
\bibfield{author}{\bibinfo{person}{Yunzhi Tan}, \bibinfo{person}{Min Zhang},
  \bibinfo{person}{Yiqun Liu}, {and} \bibinfo{person}{Shaoping Ma}.}
  \bibinfo{year}{2016}\natexlab{}.
\newblock \showarticletitle{Rating-boosted latent topics: Understanding users
  and items with ratings and reviews}. In \bibinfo{booktitle}{\emph{Proceedings
  of the Twenty-Fifth International Joint Conference on Artificial
  Intelligence}}. \bibinfo{publisher}{AAAI Press}, \bibinfo{pages}{2640--2646}.
\newblock


\bibitem[\protect\citeauthoryear{Wang, Huang, Liu, Ma, Chen, and
  Veijalainen}{Wang et~al\mbox{.}}{2016}]%
        {TOIS2016RANK}
\bibfield{author}{\bibinfo{person}{Shuaiqiang Wang}, \bibinfo{person}{Shanshan
  Huang}, \bibinfo{person}{Tie-Yan Liu}, \bibinfo{person}{Jun Ma},
  \bibinfo{person}{Zhumin Chen}, {and} \bibinfo{person}{Jari Veijalainen}.}
  \bibinfo{year}{2016}\natexlab{}.
\newblock \showarticletitle{Ranking-oriented collaborative filtering: A
  listwise approach}.
\newblock \bibinfo{journal}{\emph{ACM Transactions on Information Systems}}
  \bibinfo{volume}{35}, \bibinfo{number}{2} (\bibinfo{year}{2016}),
  \bibinfo{pages}{10:1--10:28}.
\newblock


\bibitem[\protect\citeauthoryear{Wang, He, Feng, Nie, and Chua}{Wang
  et~al\mbox{.}}{2018}]%
        {Wang2018TEM}
\bibfield{author}{\bibinfo{person}{Xiang Wang}, \bibinfo{person}{Xiangnan He},
  \bibinfo{person}{Fuli Feng}, \bibinfo{person}{Liqiang Nie}, {and}
  \bibinfo{person}{Tat-Seng Chua}.} \bibinfo{year}{2018}\natexlab{}.
\newblock \showarticletitle{TEM: Tree-enhanced Embedding Model for Explainable
  Recommendation}. In \bibinfo{booktitle}{\emph{Proceedings of the 2018 World
  Wide Web Conference}} \emph{(\bibinfo{series}{WWW '18})}.
  \bibinfo{publisher}{International World Wide Web Conferences Steering
  Committee}, \bibinfo{address}{Republic and Canton of Geneva, Switzerland},
  \bibinfo{pages}{1543--1552}.
\newblock


\bibitem[\protect\citeauthoryear{Xiao, Ye, He, Zhang, Wu, and Chua}{Xiao
  et~al\mbox{.}}{2017}]%
        {AFM2017}
\bibfield{author}{\bibinfo{person}{Jun Xiao}, \bibinfo{person}{Hao Ye},
  \bibinfo{person}{Xiangnan He}, \bibinfo{person}{Hanwang Zhang},
  \bibinfo{person}{Fei Wu}, {and} \bibinfo{person}{Tat-Seng Chua}.}
  \bibinfo{year}{2017}\natexlab{}.
\newblock \showarticletitle{Attentional factorization machines: Learning the
  weight of feature interactions via attention networks}. In
  \bibinfo{booktitle}{\emph{Proceedings of the 26th International Joint
  Conference on Artificial Intelligence}}. \bibinfo{publisher}{AAAI Press},
  \bibinfo{pages}{3119--3125}.
\newblock


\bibitem[\protect\citeauthoryear{Xu, Lam, and Lin}{Xu et~al\mbox{.}}{2014}]%
        {CMR2014}
\bibfield{author}{\bibinfo{person}{Yinqing Xu}, \bibinfo{person}{Wai Lam},
  {and} \bibinfo{person}{Tianyi Lin}.} \bibinfo{year}{2014}\natexlab{}.
\newblock \showarticletitle{Collaborative filtering incorporating review text
  and co-clusters of hidden user communities and item groups}. In
  \bibinfo{booktitle}{\emph{Proceedings of the 23rd ACM International
  Conference on Conference on Information and Knowledge Management}}. ACM,
  \bibinfo{pages}{251--260}.
\newblock


\bibitem[\protect\citeauthoryear{Yang, Hsieh, Yang, Pollak, Dell, Belongie,
  Cole, and Estrin}{Yang et~al\mbox{.}}{2017}]%
        {TOIS2017YUM}
\bibfield{author}{\bibinfo{person}{Longqi Yang}, \bibinfo{person}{Cheng-Kang
  Hsieh}, \bibinfo{person}{Hongjian Yang}, \bibinfo{person}{John~P. Pollak},
  \bibinfo{person}{Nicola Dell}, \bibinfo{person}{Serge Belongie},
  \bibinfo{person}{Curtis Cole}, {and} \bibinfo{person}{Deborah Estrin}.}
  \bibinfo{year}{2017}\natexlab{}.
\newblock \showarticletitle{Yum-Me: A personalized nutrient-based meal
  recommender system}.
\newblock \bibinfo{journal}{\emph{ACM Transactions on Information Systems}}
  \bibinfo{volume}{36}, \bibinfo{number}{1} (\bibinfo{year}{2017}),
  \bibinfo{pages}{7:1--7:31}.
\newblock


\bibitem[\protect\citeauthoryear{Zhang and Wang}{Zhang and Wang}{2016}]%
        {ITLFM2016}
\bibfield{author}{\bibinfo{person}{Wei Zhang} {and} \bibinfo{person}{Jianyong
  Wang}.} \bibinfo{year}{2016}\natexlab{}.
\newblock \showarticletitle{Integrating topic and latent factors for scalable
  personalized review-based rating prediction}.
\newblock \bibinfo{journal}{\emph{IEEE Transactions on Knowledge and Data
  Engineering}} \bibinfo{volume}{28}, \bibinfo{number}{11}
  (\bibinfo{year}{2016}), \bibinfo{pages}{3013--3027}.
\newblock


\bibitem[\protect\citeauthoryear{Zhang, Ai, Chen, and Croft}{Zhang
  et~al\mbox{.}}{2017}]%
        {zhang2017joint}
\bibfield{author}{\bibinfo{person}{Yongfeng Zhang}, \bibinfo{person}{Qingyao
  Ai}, \bibinfo{person}{Xu Chen}, {and} \bibinfo{person}{W~Bruce Croft}.}
  \bibinfo{year}{2017}\natexlab{}.
\newblock \showarticletitle{Joint representation learning for top-n
  recommendation with heterogeneous information sources}. In
  \bibinfo{booktitle}{\emph{Proceedings of the 2017 ACM on Conference on
  Information and Knowledge Management}}. ACM, \bibinfo{pages}{1449--1458}.
\newblock


\bibitem[\protect\citeauthoryear{Zhang, Lai, Zhang, Zhang, Liu, and Ma}{Zhang
  et~al\mbox{.}}{2014a}]%
        {Zhang:2014:EFM:2600428.2609579}
\bibfield{author}{\bibinfo{person}{Yongfeng Zhang}, \bibinfo{person}{Guokun
  Lai}, \bibinfo{person}{Min Zhang}, \bibinfo{person}{Yi Zhang},
  \bibinfo{person}{Yiqun Liu}, {and} \bibinfo{person}{Shaoping Ma}.}
  \bibinfo{year}{2014}\natexlab{a}.
\newblock \showarticletitle{Explicit factor models for explainable
  recommendation based on phrase-level sentiment analysis}. In
  \bibinfo{booktitle}{\emph{Proceedings of the 37th international ACM SIGIR
  Conference on Research and Development in Information Retrieval}}. ACM,
  \bibinfo{pages}{83--92}.
\newblock


\bibitem[\protect\citeauthoryear{Zhang, Zhang, Zhang, Liu, and Ma}{Zhang
  et~al\mbox{.}}{2014b}]%
        {zhang2014users}
\bibfield{author}{\bibinfo{person}{Yongfeng Zhang}, \bibinfo{person}{Haochen
  Zhang}, \bibinfo{person}{Min Zhang}, \bibinfo{person}{Yiqun Liu}, {and}
  \bibinfo{person}{Shaoping Ma}.} \bibinfo{year}{2014}\natexlab{b}.
\newblock \showarticletitle{Do users rate or review?: Boost phrase-level
  sentiment labeling with review-level sentiment classification}. In
  \bibinfo{booktitle}{\emph{Proceedings of the 37th International ACM SIGIR
  Conference on Research and Development in Information Retrieval}}. ACM,
  \bibinfo{pages}{1027--1030}.
\newblock


\bibitem[\protect\citeauthoryear{Zheng, Noroozi, and Yu}{Zheng
  et~al\mbox{.}}{2017}]%
        {Zheng:2017:JDM:3018661.3018665}
\bibfield{author}{\bibinfo{person}{Lei Zheng}, \bibinfo{person}{Vahid Noroozi},
  {and} \bibinfo{person}{Philip~S. Yu}.} \bibinfo{year}{2017}\natexlab{}.
\newblock \showarticletitle{Joint deep modeling of users and items using
  reviews for recommendation}. In \bibinfo{booktitle}{\emph{Proceedings of the
  Tenth ACM International Conference on Web Search and Data Mining}}. ACM,
  \bibinfo{pages}{425--434}.
\newblock


\end{thebibliography}

\end{document}